\documentclass[twocolumn,aps,prd,epsf,showpacs]{revtex4}

\usepackage[dvips]{graphicx}
\usepackage{epsfig}
\usepackage{color}
\usepackage{amsmath}
\usepackage[utf8]{inputenc}
\usepackage[english]{babel}
\usepackage{tikz}
\usepackage[normalem]{ulem} 
\usepackage{braket} 

\newcommand{\be}{\begin{equation}}
\newcommand{\ee}{\end{equation}}
\newcommand{\bea}{\begin{eqnarray}}
\newcommand{\eea}{\end{eqnarray}}
\newcommand{\bi}{\begin{itemize}}
\newcommand{\ie}{\item}
\newcommand{\ei}{\end{itemize}}
\newcommand{\non}{\nonumber}

\newcommand{\ltapprox}{\raisebox{-0.5ex}{$\,\stackrel{<}{\scriptstyle\sim}\,$}}

\begin{document}


\title{Evidence for the existence of $u d \bar{b} \bar{b}$ and the non-existence of $s s \bar{b} \bar{b}$ and $c c \bar{b} \bar{b}$ tetraquarks from lattice QCD}

\author{$^{(1)}$Pedro Bicudo}

\author{$^{(2,3)}$Krzysztof Cichy}

\author{$^{(2)}$Antje Peters}

\author{$^{(2)}$Bj\"orn Wagenbach}

\author{$^{(2)}$Marc Wagner}

\affiliation{\vspace{0.1cm}$^{(1)}$CFTP, Dep.\ F\'{\i}sica, Instituto Superior T\'ecnico, Universidade de Lisboa, Av.\ Rovisco Pais, 1049-001 Lisboa, Portugal}

\affiliation{\vspace{0.1cm}$^{(2)}$Johann Wolfgang Goethe-Universit\"at Frankfurt am Main, Institut f\"ur Theoretische Physik, Max-von-Laue-Stra{\ss}e 1, D-60438 Frankfurt am Main, Germany}

\affiliation{\vspace{0.1cm}$^{(3)}$Adam Mickiewicz University, Faculty of Physics, Umultowska 85, 61-614 Poznan, Poland}

\begin{abstract}
We combine lattice QCD results for the potential of two static antiquarks in the presence of two quarks
$q q$ of finite mass and quark model techniques to study possibly existing $q q \bar{b} \bar{b}$
tetraquarks. While there is strong indication for a bound four-quark state for $q q = (ud-du) /
\sqrt{2}$, i.e.\ isospin $I=0$, we find clear evidence against the existence of corresponding tetraquarks
with $q q \in \{ uu , (ud+du) / \sqrt{2} , dd \}$, i.e.\ isospin $I=1$, $q q = s s$ and $q q = c c$.
\end{abstract}


\maketitle


\section{Introduction}

Exotic hadrons have been proposed many years ago. As soon as quarks were found in the sixties, it became
clear that systems more complex than standard mesons ($q \bar{q}$ states) and baryons ($q q q$ states)
could possibly exist. However, exotic hadrons are very elusive systems. Confirming their existence or
non-existence still remains one of the main challenges of particle physics. 

Frequently discussed exotic hadrons are tetraquarks \cite{Jaffe:1976ig,Jaffe:2004ph}, which are
four-quark bound states composed of two quarks and two antiquarks. There are several hadronic resonances
which are tetraquark candidates. Among them are the light scalar mesons $\sigma$, $\kappa$, $f_0(980)$
and $a_0(980)$ as well as the heavier mesons $D_{s0}^\ast$ and $D_{s1}$. However, these systems have
quantum numbers also consistent with a standard $q \bar{q}$ structure and their masses are not too
different from what is expected in a $q \bar{q}$ picture. Thus, it is hard to rigorously argue that they
are indeed predominantly tetraquarks. On the other hand, there are also candidates which have quantum
numbers or masses typical for tetraquarks, but not for standard $q \bar{q}$ mesons. For example $\pi_1^{-
+}$ has exotic quantum numbers $J^{P C} = 1^{- +}$ or $Z_c^\pm$ and $Z_b^\pm$ masses and decay products
strongly suggest hidden $c \bar{c}$ or $b \bar{b}$ pairs, while their electrical charge $\pm 1$ indicates
isospin $I = 1$. While the evidence for $\pi_1^{- +}$ is not conclusive and the $Z_b^\pm$ claimed by
the BELLE collaboration \cite{Belle:2011aa} remains to be confirmed by different experimental
collaborations, the $Z_c^\pm$ has received a series of experimental observations by the BELLE
collaboration \cite{Liu:2013dau,Chilikin:2014bkk}, the Cleo-C collaboration \cite{Xiao:2013iha}, the
BESIII collaboration
\cite{Ablikim:2013mio,Ablikim:2013emm,Ablikim:2013wzq,Ablikim:2013xfr,Ablikim:2014dxl} and the LHCb
collaboration \cite{Aaij:2014jqa}. Nevertheless, the $Z_c^\pm$ would profit by more comprehensive
measurements of its decay channels. We expect the existing and future experimental collaborations to
continue the study of present tetraquark candidates and to possibly also discover further ones.

The theoretical study of tetraquarks is crucial to confirm and correctly interpret corresponding
experimental observations and could as well provide information in which channels tetraquarks may be
found. However, tetraquark studies face a number of difficulties, e.g.\ (1)~tetraquarks are usually open
to meson-meson decay, (2)~tetraquarks are complex relativistic four-body systems, (3)~quark models still
fail to reproduce sectors of standard hadronic spectra and, thus, are not yet sufficiently well
calibrated to reliably predict tetraquarks. 

In this work, we study the existence/non-existence of tetraquarks with two heavy bottom antiquarks
$\bar{b} \bar{b}$. To this end, we use potentials of two static antiquarks in the presence of two quarks
$q q$ of finite mass, which we compute using lattice QCD. We extend recent studies of $q q \bar{b}
\bar{b}$ tetraquarks \cite{Bicudo:2012qt,Brown:2012tm}, where $q q \in \{ (ud-du)/\sqrt{2} \ , \ uu ,
(ud+du)/\sqrt{2} , dd \}$, to similar systems with heavier quarks, $q q = s s$ and $q q = c c$. In the
future, we also plan to extend our investigations to the $b \bar{b}$ tetraquarks claimed by the BELLE
Collaboration \cite{Belle:2011aa}. Such tetraquarks with a $b \bar{b}$ pair are, however, rather
difficult to study with lattice QCD, since they couple to several decay channels.

We avoid some of the technical difficulties of studying tetraquarks following a strategy already
identified in the eighties \cite{Ballot:1983iv}. We search for bound states rather than for resonances,
to avoid open decay channels. Moreover, by using $\bar{b} \bar{b}$ potentials obtained by lattice QCD
computations, we largely avoid the calibration problem of quark models. Very heavy antiquarks such as
$\bar{b}$ allow for the Born-Oppenheimer approximation \cite{Born:1927}. For the two lighter quarks $q
q$, the heavy antiquarks $\bar{b} \bar{b}$ can be approximated as static color charges, which allows to
determine the light quark energy using lattice QCD. On the other hand, once the energy of the light
quarks $q q$ is determined, it can be utilized as an effective potential for the heavy antiquarks
$\bar{b} \bar{b}$.

Our lattice QCD computation goes beyond computations with four static quarks, which show a clear
evidence for four-body tetraquark potentials \cite{Alexandrou:2004ak,Okiharu:2004ve} and tetraquark flux
tubes \cite{Cardoso:2011fq,Cardoso:2012uka}. On the other hand, lattice QCD computations with four quarks
of finite mass are extremely difficult and have found neither evidence for charmed tetraquark bound
states with $\bar u \bar d c c$ flavor \cite{Guerrieri:2014nxa} nor for resonances in the $Z_c$ family
\cite{Prelovsek:2014swa}.

This paper is organized as follows. In section~\ref{sec:II_model}, we briefly review the quark model and
discuss qualitative expectations regarding $q q \bar{b} \bar{b}$ four-quark systems. In
section~\ref{sec:III_latticefit}, we discuss the lattice QCD computation of $\bar{b} \bar{b}$ potentials
in the presence of two lighter quarks $q q$ and provide parameterizations of these results by continuous
functions. In section~\ref{sec:IV_binding}, we use these parameterizations in model calculations and
check for the existence of bound states, which would indicate the existence of tetraquarks. We conclude
in section~\ref{sec:V_conclusion}.


\section{\label{sec:II_model}Modeling the $\bar{b} \bar{b}$ interaction in the presence of two light quarks $q q$}

In the following, we discuss quark model expectations regarding the qualitative behavior of a $q q
\bar{b} \bar{b}$ four-quark systems, where $q$ denotes either a light $u$, $d$, $s$ or $c$ quark
\footnote{In the context of this paper, a light quark $q$ is a quark significantly lighter than a $b$
quark, i.e.\ $q \in \{ u,d,s,c \}$.}. In particular, we are interested in the $\bar{b} \bar{b}$
interaction in the presence of two light quarks $q q$. The qualitative expectations are confirmed by
corresponding lattice QCD results, which are discussed in section~\ref{sec:III_latticefit}. The main
purpose of these model considerations is to motivate a suitable fit function for the lattice QCD $\bar{b}
\bar{b}$ potential results, which is used in section~\ref{sec:IV_binding} in the Schr\"odinger equation
to check whether and in which channels bound four-quark states, i.e.\ tetraquarks, exist.


\subsection{\label{SEC500}The quark-antiquark / quark-quark potential at small separations}

In the original quark model \cite{De Rujula:1975ge}, the quark-antiquark and the quark-quark (or
equivalently anti\-quark-antiquark) potentials at small separations $r = |\mathbf{r}_i - \mathbf{r}_j|$
are dominated by one-gluon exchange similar to the Fermi-Breit interaction,
\bea
\nonumber & & \hspace{-0.7cm} V_{i j}(\mathbf{r}_i,\mathbf{s}_i,\mathbf{r}_j,\mathbf{s}_j) \ \ = \ \ -\frac{C \alpha_s}{4}
\non \\
\label{eq:fermibreit} & & \bigg(\frac{1}{r} - \frac{\pi}{2} \delta^3(\mathbf{r}) \left(\frac{1}{{m_i}^2} + \frac{1}{{m_j}^2} + \frac{16 \mathbf{s}_i \cdot \mathbf{s}_j}{3 m_i m_j}\right) + \ldots\bigg)
\eea
($i,j$ are the (anti)quark indices, $\mathbf{r}_i$, $\mathbf{s}_i$ and $m_i$ denote their positions, spins and masses, respectively). Since we are exclusively interested in ground states, we have specialized eq.\ (\ref{eq:fermibreit}) to angular momentum $l=0$. The quark model has been improved (cf.\ e.g.\ \cite{Godfrey:1985xj,Capstick:1986bm}), but maintains its main ingredients. $C$ depends on the color orientation of the (anti)quarks, which can be specified by a $3 \times 3$ matrix $\Lambda$. For a quark-antiquark pair $\bar{q}_i \Lambda q_j$
\be
C \ \ = \ \ +\sum_a \textrm{Tr}\Big(\lambda^a \Lambda \lambda^a \Lambda^\dagger\Big) ,
\ee
while for a quark-quark pair $q_i{}^T \Lambda q_j$
\be
C \ \ = \ \ -\sum_a \textrm{Tr}\Big(\lambda^a \Lambda \lambda^a{}^T \Lambda^\dagger\Big)
\ee
with the Gell-Mann matrices $\lambda^a$, $a=1,\ldots,8$. For example, $\Lambda_{A B} = \delta_{A B} /
\sqrt{3}$ describes the $q \bar{q}$ color singlet, while $\Lambda_{A B} = \epsilon_{A B 3} / \sqrt{2}$ is
one of three independent possibilities to realize a $q q$ color triplet. In Table~\ref{ta:casimirs}, the
resulting values for $C$ for the $q \bar{q}$ singlet and octet and the $q q$ triplet and sextet color
orientations are listed.

\begin{table}[htb]
\begin{tabular}{c|c|c|c|c}
\hline
color & $\ $$q \bar{q}$ singlet$\ $ & $\ $$q \bar{q}$ octet$\ $ & $\ $$q q$ triplet$\ $ & $\ $$q q$ sextet$\ $ \\
orientation & $1$ & $8$ & $\bar{3} \textrm{ (and } 3 \textrm{)}$ & $\bar{6} \textrm{ (and } 6 \textrm{)}$ \\
\hline
$C$ & $+16/3$ & $-2/3$ & $+8/3$ & $-4/3$ \\
 & (attractive) & (repulsive) & (attractive) & (repulsive) \\
\hline
\end{tabular}
\caption{\label{ta:casimirs}The color factors $C$ for the $q \bar{q}$ singlet and octet and the $q q$
triplet and sextet color orientations.}
\end{table}

Lattice QCD confirms that the static color singlet potential at small separations $r$ can be described
reasonably well by one-gluon-exchange (cf.\ e.g.\ \cite{Brambilla:2010pp,Jansen:2011vv}, where a matching
of lattice QCD and perturbative results is done). At larger separations, it becomes linear with certain
$1/r$-corrections due to string vibrations \cite{Luscher:2002qv}. One can crudely estimate $\alpha_s$
appearing in (\ref{eq:fermibreit}) by considering the color singlet $q \bar{q}$. In that case $C =
+16/3$, while string vibrations lead to $V_{i j} \approx -\pi/12 r$ at intermediate separations,
resulting in $\alpha_s \approx \pi/16$. While this estimate is most appropriate for static quarks,
$\alpha_s$ is expected to be somewhat larger for quarks of finite mass
\cite{Godfrey:1985xj,Capstick:1986bm}.

The only spin dependent term in (\ref{eq:fermibreit}) is the hyperfine interaction proportional to
$\mathbf s_i \cdot \mathbf s_j $, which is pathological in the original quark model due to the Dirac
delta (cf.\ eq.\ (\ref{eq:fermibreit})). In the relativistic quark model, however, this interaction is
smoother and, hence, well behaved \cite{Godfrey:1985xj,Capstick:1986bm}. Clearly, the interaction is
weaker for a spin triplet than for a spin singlet.

To summarize, whether the potential between a quark and another quark or antiquark is attractive or
repulsive depends on their color orientation. For small separations it is approximately Coulomb-like with
the color factors $C$ collected in Table~\ref{ta:casimirs}. The hyperfine term enhances the interaction
for a spin singlet and decreases it for a spin triplet.


\subsection{\label{SEC507}Qualitative discussion of the $q q \bar{b} \bar{b}$ system}

For the particular case of the $q q \bar{b} \bar{b}$ system, where the $\bar{b} \bar{b}$ pair is
significantly heavier than the light $q q$ pair, we utilize the Born-Oppenheimer approximation
\cite{Born:1927}: for the light quarks, the heavy antiquarks can be regarded as static color charges;
once the energy of the light quarks is determined, it can be used as an effective potential for the heavy
antiquarks. We assume that at small $\bar{b} \bar{b}$ separations $r$, the $\bar{b}$ quarks interact
according to the quark model discussed in section~\ref{SEC500}, while at larger separations their
interaction is screened by the light quarks, i.e.\ the four quarks form two rather weakly interacting
$B_{(s,c)}^{(\ast)}$ mesons ($B_{(s,c)}^{(\ast)}$ denotes either a $B$, $B^\ast$, $B_s$, $B_s^\ast$,
$B_c$ or $B_c^\ast$ meson).


\subsubsection*{Expectations for the $\bar{b} \bar{b}$ interaction at small separations $r$}

\begin{figure}[htb]
\includegraphics[width=1.0\columnwidth]{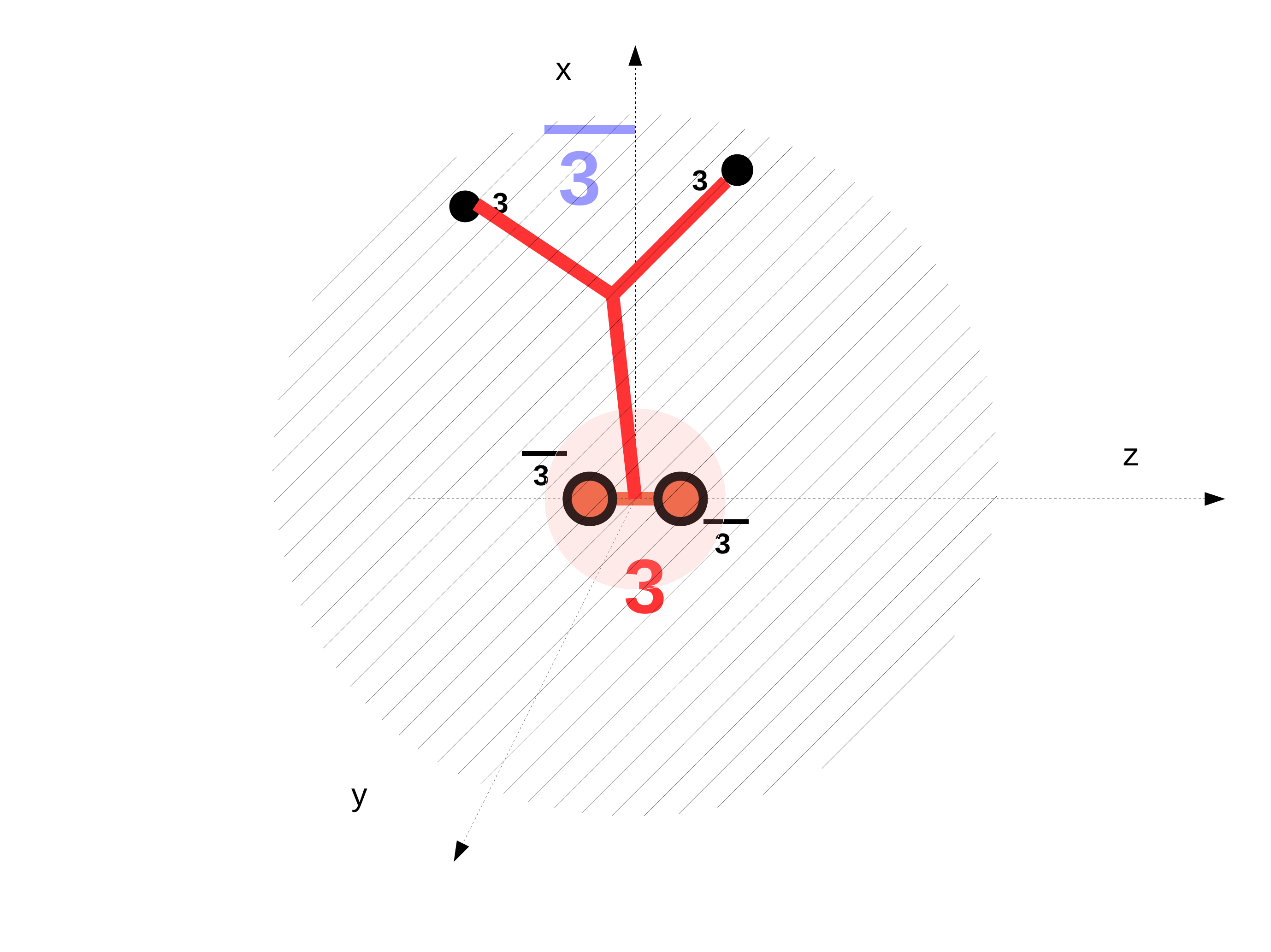}
\caption{\label{fig:001}(Color online) At small $\bar{b} \bar{b}$ separations $r$, the heavy antiquarks
$\bar{b} \bar{b}$ form an antidiquark, which corresponds to a color triplet. There is essentially no
screening of the $\bar{b} \bar{b}$ interaction due to the much farther separated light quarks $q q$.}
\end{figure}

\bi
\ie The spin interaction of the $\bar{b}$ quarks is quite small and can possibly be neglected, since it
is proportional to $1 / m_b{}^2$ (cf.\ eq.\ (\ref{eq:fermibreit})).

\ie In case of a bound $q q \bar{b} \bar{b}$ state, i.e.\ a tetraquark, the antiquarks $\bar{b} \bar{b}$
are expected to be in a color triplet $3$, which is attractive, and not in a color sextet $6$, which is
repulsive (cf.\ also Table~\ref{ta:casimirs}). In other words, at small separations $r$, the antiquarks
$\bar{b} \bar{b}$ form an antidiquark as depicted in Figure~\ref{fig:001}.

\ie Because the complete four quark system $q q \bar{b} \bar{b}$ necessarily forms a color singlet, the
light quarks $q q$ must be in a color antitriplet $\bar{3}$.

\ie Since this color antitriplet is antisymmetric, and since the light quarks $q q$ are assumed to be in
a spatially symmetric s-wave, the Pauli principle implies a symmetric spin-flavor structure. This can
either be a spin singlet with an antisymmetric flavor combination or a spin triplet with a symmetric
flavor combination. Indeed, when studying light $u$ and $d$ quarks in the presence of two static
antiquarks using lattice QCD, two attractive channels have been found
\cite{Wagner:2010ad,Wagner:2011ev,Bicudo:2012qt}. As expected, these are a (spin) \textit{scalar
isosinglet} ($j=0$, $I=0$, where $j$ denotes the spin of the light quarks $q q$) and a (spin)
\textit{vector isotriplet} ($j=1$, $I=1$). The scalar isosinglet is more attractive, as expected from the
hyperfine interaction in eq.\ (\ref{eq:fermibreit}), i.e.\ the lattice QCD results confirm the
qualitative quark model expectations.

\ie When studying two identical light quarks $q q = s s$ or $q q = c c$, which are symmetric in flavor,
the only attractive channel is a spin triplet. However, it is conceptually interesting to consider two
hypothetical degenerate flavors with the mass of strange or charm quarks and then also investigate spin
singlets with flavor structure $q q = (s^{(1)}s^{(2)}-s^{(2)}s^{(1)})/\sqrt{2}$ and $q q =
(c^{(1)}c^{(2)}-c^{(2)}c^{(1)})/\sqrt{2}$.
\ei


\subsubsection*{Expectations for the $\bar{b} \bar{b}$ interaction at large separations $r$}

\begin{figure}[htb]
\includegraphics[width=1.0\columnwidth]{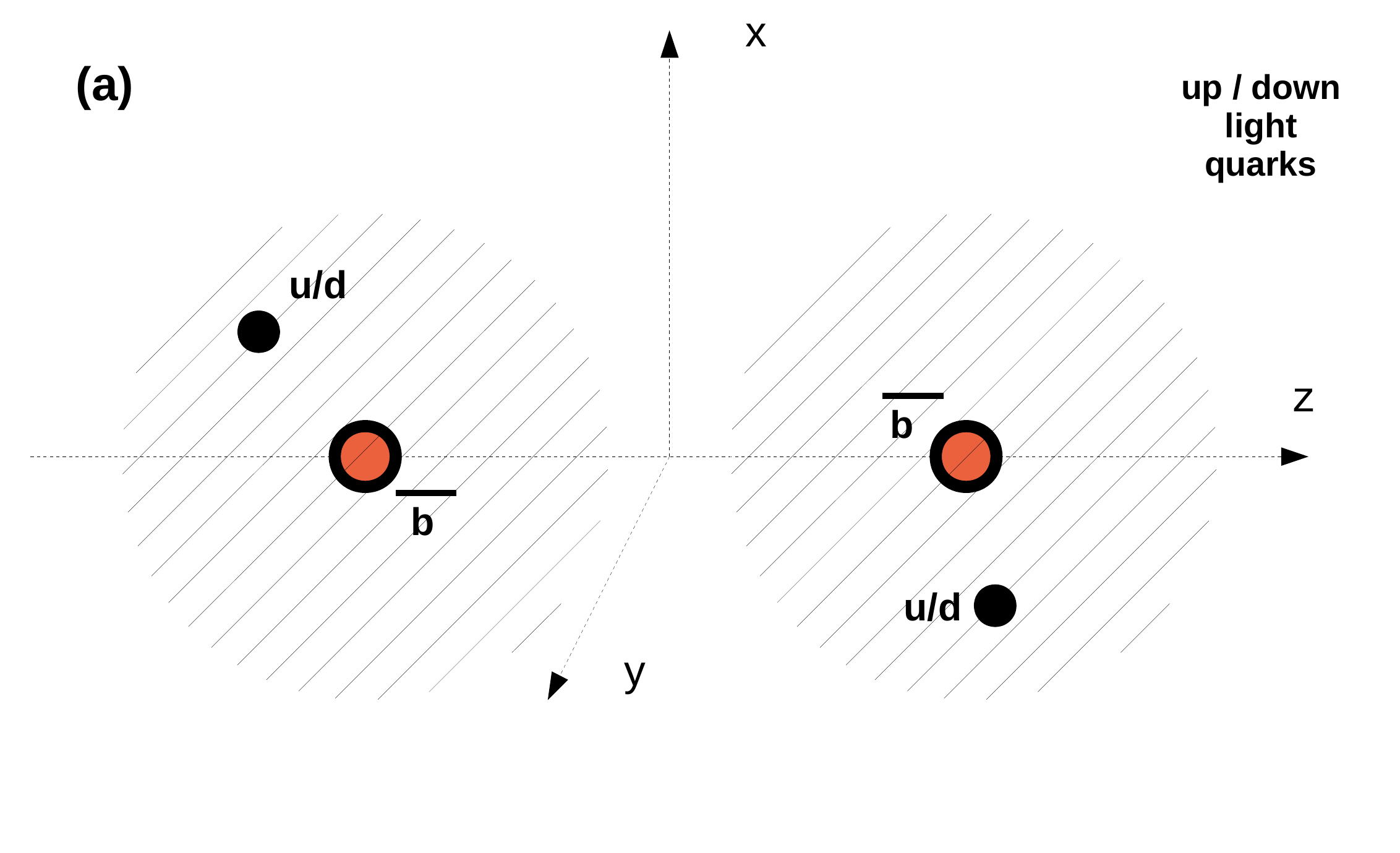}
\includegraphics[width=1.0\columnwidth]{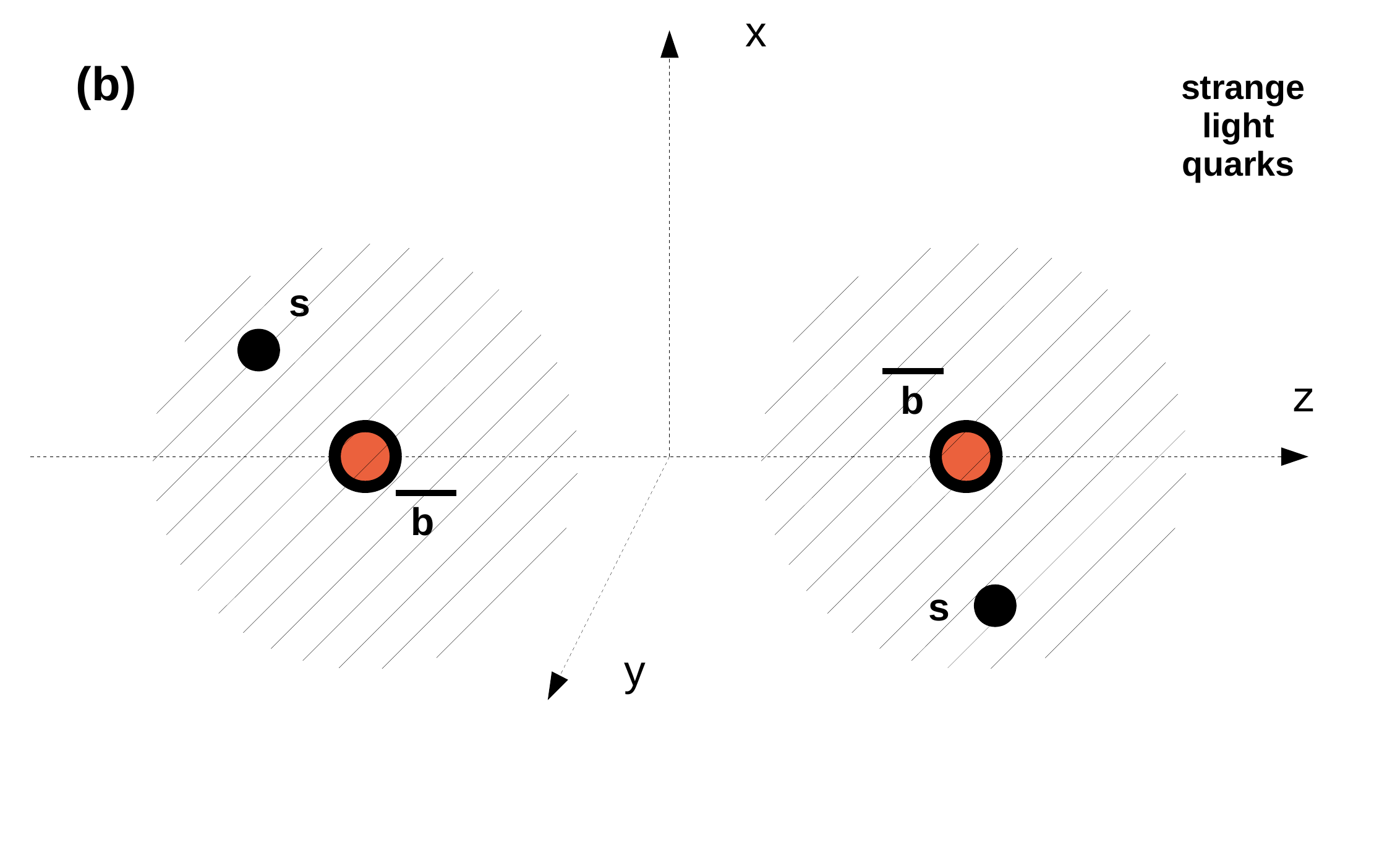}
\includegraphics[width=1.0\columnwidth]{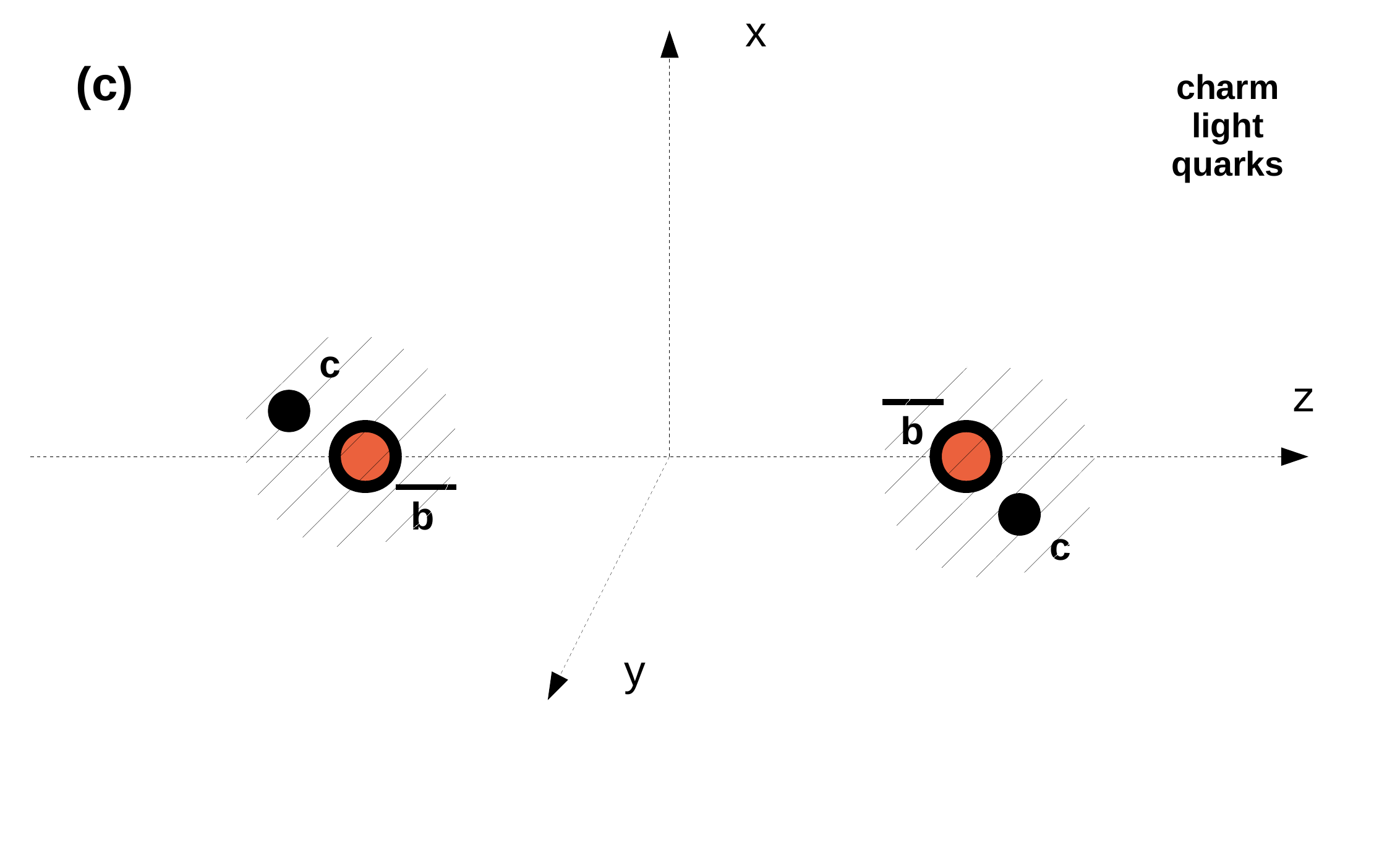}
\caption{\label{fig:002}(Color online). At large $\bar{b} \bar{b}$ separations $r$, the $q q \bar{b}
\bar{b}$ system is essentially a system of two $B_{(s,c)}^{(\ast)}$ mesons. The color charge of each of
the antiquarks $\bar{b}$ is almost completely screened by one of the light quarks $q$. (a)~$q \in
\{u,d\}$. (b)~$q=s$. (c)~$q=c$.}
\end{figure}

\bi
\ie At large separations $r$, screening of the $\bar{b} \bar{b}$ interaction is expected due to the light
quarks $q q$, as illustrated in Figure~\ref{fig:002}. When the $\bar{b} \bar{b}$ separation is larger
than around two times the radius of a $B_{(s,c)}^{(\ast)}$ meson, there is essentially no overlap between
the wave functions of the light quarks and, consequently, the $\bar{b} \bar{b}$ interaction practically
vanishes.

\ie The more massive the light quarks are, the more compact their wave functions in the
$B_{(s,c)}^{(\ast)}$ mesons, as shown in Figure~\ref{fig:002}(a), (b) and (c) and, thus, the stronger the
screening. In other words, the corresponding $\bar{b} \bar{b}$ potential becomes more and more narrow and
will at some point not anymore be able to host a bound state. Consequently, for a sufficiently heavy pair
of light quarks $q q$ the screening should prevent the formation of $q q \bar{b} \bar{b}$ tetraquarks.
\ei


\subsubsection*{Quantum numbers of possibly existing $q q \bar{b} \bar{b}$ tetraquarks}
We study exclusively states which correspond for large $\bar{b} \bar{b}$ separations to pairs of
$B_{(s,c)}^{(\ast)}$ mesons in a spatially symmetric s-wave. Therefore, the parity of these states is
positive, i.e.\ $P = +$ (the product of the parity quantum numbers of the two mesons, which are both
negative).

As argued above, the two antiquarks $\bar{b} \bar{b}$ are expected to be in an antisymmetric color
triplet. Since their flavor is symmetric, their spin $j_b$ must also be symmetric due to the Pauli
principle, i.e.\ $j_b = 1$. Similarly, for an antisymmetric $q q$ flavor combination, i.e.\ $qq = (ud -
du)/\sqrt{2}$, $j = 0$, while for symmetric flavor combinations, i.e.\ $qq \in \{ uu , (ud + du)/\sqrt{2}
, dd \ , \ ss \ , \ cc \}$, $j = 1$. The total spin $J$ of the $q q \bar{b} \bar{b}$ system is the
combination of $j$ and $j_b$.

Altogether, the possibly existing $q q \bar{b} \bar{b}$ tetraquarks we are going to investigate have the
following quantum numbers:
\bi
\ie $qq = (ud - du)/\sqrt{2}$: \\ $I(J^P) = 0(1^+)$.

\ie $qq \in \{ uu , (ud + du)/\sqrt{2} , dd \}$: \\ $I(J^P) \in \{ 1(0^+) , 1(1^+) , 1(2^+) \}$.

\ie $qq \in \{ ss , cc \}$: \\ $I(J^P) \in \{ 0(0^+) , 0(1^+) , 0(2^+) \}$.
\ei


\subsection{\label{SEC388}Fit function for lattice QCD $\bar{b} \bar{b}$ potential results}

Using lattice QCD, one can compute $\bar{b} \bar{b}$ potentials in the static limit (i.e.\ for $m_b
\rightarrow \infty$) from first principles, i.e.\ from the QCD Lagrangian (cf.\
\cite{Wagner:2010ad,Wagner:2011ev} and section~\ref{sec:III_latticefit}). Of course, these potentials can
be obtained only for a limited number of discrete separations $r$. Therefore, a suitable fit function is
required, to interpolate between the lattice QCD results and also to extrapolate beyond them. This fit
function is based on the qualitative expectations discussed above and will be used in the Schr\"odinger
equation in section~\ref{sec:IV_binding}, where we determine whether and in which channels bound
four-quark states exist.

For two heavy antiquarks $\bar{b} \bar{b}$ inside a cloud of two light quarks $q q$, i.e.\ at small
$\bar{b} \bar{b}$ separations, we expect a Coulomb-like potential of order $-2 \alpha_s / 3 r \approx -
\pi / 24 r$ corresponding to a color triplet. At larger separations $r$, the potential will be screened
by the light quarks $q q$. This is due to the decrease of the wave function $\psi$ of each of the light
quarks with respect to their separations from the heavy antiquarks. One expects this decrease to follow
an exponential of a power of $r$, i.e.\ $\psi \propto \exp(-(r/d)^p)$, where $d$ roughly describes the
size of each of the $\bar{b} q$ systems, i.e.\ the size of a $B_{(s,c)}^{(\ast)}$ meson $\ltapprox 0.5 \,
\textrm{fm}$. The parameter $p$ characterizes the radial profile of the light quark wave function inside
the $B_{(s,c)}^{(\ast)}$ meson. Assuming the $q \bar{b}$ interaction inside the $B_{(s,c)}^{(\ast)}$
meson is dominated by a linear confining potential, one can estimate the parameter $p$. In the case,
where the quark $q$ is rather heavy, e.g.\ $q = c$, the corresponding non-relativistic Schr\"odinger
equation is solved by Airy functions, resulting in $p = 3/2$. A similar but relativistic treatment for
a lighter quark yields $p = 2$ instead.

These considerations suggest the following fit function for lattice QCD $\bar{b} \bar{b}$ potential
results:
\be
\label{eq:ansatz} V(r) \ \ = \ \ -\frac{\alpha}{r} \exp\left(-\left(\frac{r}{d}\right)^p\right) + V_0 ,
\ee
where it is expected that $\alpha \approx 2 \alpha_s / 3 \approx \pi / 24 \approx 0.13$, $d \ltapprox
0.5 \, \textrm{fm}$ and $p \approx 1.5 \ldots 2.0$. The constant $V_0$ is necessary to account for twice
the mass of the static-light meson. As will be demonstrated in the following section, this fit function
is consistent with lattice QCD results and the crude quantitative expectations for $\alpha$, $d$ and $p$
are fulfilled.


\section{\label{sec:III_latticefit}Lattice QCD computation of the $\bar{b} \bar{b}$ interaction in the presence of two light quarks $q q$}

To determine the effective $\bar{b} \bar{b}$ potential quantitatively, we use lattice QCD and consider
the limit of infinitely heavy $\bar{b}$ quarks, i.e.\ the static limit. The first lattice computations of
such potentials have been performed in the quenched approximation (cf.\ e.g.\
\cite{Stewart:1998hk,Michael:1999nq,Cook:2002am,Doi:2006kx,Detmold:2007wk}). Recently, also computations
with dynamical sea quarks have been performed
\cite{Wagner:2010ad,Bali:2010xa,Wagner:2011ev,Bicudo:2012qt,Brown:2012tm,Wagenbach:2014oxa}. In this
work, we extend our previous computations for light quark combinations $q q \in \{ (ud-du)/\sqrt{2} \ , \
uu , (ud+du)/\sqrt{2} , dd \}$ \cite{Wagner:2010ad,Wagner:2011ev} by similar computations with strange
and charm quarks, i.e.\ $q q \in \{ (s^{(1)}s^{(2)}-s^{(2)}s^{(1)})/\sqrt{2} \ , \ ss \ , \
(c^{(1)}c^{(2)}-c^{(2)}c^{(1)})/\sqrt{2} \ , \ cc \}$.


\subsection{Lattice QCD setup}

We have performed computations using two ensembles of gauge link configurations generated by the
European Twisted Mass Collaboration (ETMC) with 2 dynamical quark flavors. The quark action is Wilson
twisted mass tuned to maximal twist, while the gluon action is tree-level Symanzik improved. Most
importantly, this guarantees automatic $\mathcal{O}(a)$ improvement of spectral quantities, i.e.\
discretization errors in the resulting $\bar{b} \bar{b}$ potentials appear only quadratically in the
lattice spacing $a$. Information about these ensembles is collected in Table~\ref{ta:ensembles}. Further
details, in particular regarding their generation, can be found in \cite{Boucaud:2008xu,Baron:2009wt}.

\begin{table}[htb]
\begin{tabular}{c|c|c|c|c|c}
\hline
$\beta$ & size & $\mu_l$ & $a$ in fm & $m_\pi$ in MeV & configurations \\
\hline
$3.90$ & $24^3 \times 48$ & $0.0040\phantom{0}$ & $0.079$ & $340$ & $480$ \\
$4.35$ & $32^3 \times 64$ & $0.00175$           & $0.042$ & $352$ & $100$ \\
\hline
\end{tabular}
\caption{\label{ta:ensembles}Ensembles of gauge link configurations used for the computation of $\bar{b}
\bar{b}$ potentials ($\beta$: inverse gauge coupling; size: number of lattice sites; $\mu_l$: bare $u/d$
quark mass in lattice units; $a$: lattice spacing; $m_\pi$: pion mass; configurations: number of gauge
link configurations used).}
\end{table}

For $\bar{b} \bar{b}$ potentials in the presence of two light quarks $q q$ with $q \in \{ u , d \}$, we
reuse our lattice QCD results from \cite{Wagner:2010ad,Wagner:2011ev}, which were obtained using the
ensemble with the coarser lattice spacing $a \approx 0.079 \, \textrm{fm}$. For $q \in \{ s , c \}$, the
$\bar{b} \bar{b}$ interaction is screened at significantly smaller $\bar{b} \bar{b}$ separations (cf.\
the discussion in section~\ref{SEC507} and Figure~\ref{fig:002}). To be able to resolve the corresponding
potentials properly, we decided to use for flavor combinations $q q \in \{
(s^{(1)}s^{(2)}-s^{(2)}s^{(1)})/\sqrt{2} \ , \ ss \ , \ (c^{(1)}c^{(2)}-c^{(2)}c^{(1)})/\sqrt{2} \ , \ cc
\}$ another ensemble with a finer lattice spacing $a \approx 0.042 \, \textrm{fm}$.
Although the physical extent of the lattice for this ensemble is much smaller than for the other one,
this should not introduce significant finite volume effects at the rather small separations we are
interested in.

Note that for both ensembles, the $u/d$ quarks are unphysically heavy, corresponding to a pion mass
$m_\pi \approx 340 \, \textrm{MeV}$. Moreover, there are no $s$ and $c$ sea quarks, i.e.\ our lattice QCD
results are obtained in a partially quenched approximation. For the computation of $\bar{b} \bar{b}$
potentials in the presence of light $s$ and $c$ quarks, we also use a much smaller number of gauge link
configurations. The reason is that the propagators of the heavier $s$ and $c$ quarks introduce less
statistical noise than those for lighter $u/d$ quarks.


\subsection{\label{SEC854}Lattice QCD computation of $\bar{b} \bar{b}$ potentials}

We determine $\bar{b} \bar{b}$ potentials in the presence of two light quarks $q q$ from the exponential
decay of temporal correlation functions,
\be
\label{eq:correlationfunction} C(t,|\mathbf{r}_1 - \mathbf{r}_2|) \ \ = \ \ \bra{\Omega} \mathcal O^\dagger(t) \mathcal O(0) \ket{\Omega}
\ee
of four-quark creation operators
\begin{eqnarray}
\nonumber & & \hspace{-0.7cm} \mathcal O(t) \ \ = \ \ (\mathcal{C} \Gamma)_{AB} (\mathcal{C} \tilde{\Gamma})_{CD} \\
\label{eq:operators} & & \Big(\bar{Q}_C(\mathbf{r}_1) q_A^{(1)}(\mathbf{r}_1)\Big) \Big(\bar{Q}_D(\mathbf{r}_2) q_B^{(2)}(\mathbf{r}_2)\Big)
\end{eqnarray}
at sufficiently large $t_\textrm{min} \leq t \leq t_\textrm{max}$. Here $\bar{Q}$ denotes a static
antiquark operator approximating a $\bar{b}$ quark, $q$ is a light quark operator, $A,B,C,D$ are spin
indices, $(1),(2)$ are flavor indices and $\mathcal{C} = \gamma_0 \gamma_2$ is the charge conjugation
matrix. For the static antiquarks, the only relevant variable is their separation. Their spin components
can be combined with $\tilde{\Gamma} \in \{ (1 - \gamma_0) \gamma_5 , (1 - \gamma_0) \gamma_j \}$, $j =
1,2,3$, where the resulting $\bar{b} \bar{b}$ potential does not depend on which $\tilde{\Gamma}$ matrix
is chosen. The spin components of the two light quarks can be coupled in $16$ independent ways via
$\Gamma$, which should be an appropriately chosen combination of $\gamma$ matrices to realize definite
quantum numbers $|j_z|$ (angular momentum with respect to the axis of separation), $P$ (parity) and $P_x$
(behavior under reflections across an axis perpendicular to the axis of separation). For a more detailed
discussion of symmetries and quantum numbers, cf.\ \cite{Wagner:2010ad}.

Note that the creation operators (\ref{eq:operators}), when applied to the vacuum $| \Omega \rangle$, do
not only generate definite quantum numbers $(|j_z|,P,P_x)$, but also a structure resembling two
$B_{(s,c)}^{(\ast)}$ mesons separated by $r = |\mathbf{r}_1 - \mathbf{r}_2|$. Such operators should be
well suited to excite the ground state of the corresponding $(|j_z|,P,P_x)$ sector, in particular for
large $\bar{Q} \bar{Q}$ separations $r$, where one expects two weakly interacting $B_{(s,c)}^{(\ast)}$
mesons (cf.\ the discussion in section~\ref{SEC507}). Note, however, that the arrangement of the four
quarks $q q \bar{Q} \bar{Q}$ in the ground state is decided by QCD dynamics, i.e.\ automatically realized
in the lattice result according to QCD and not by the structure of the employed creation operators. For
example, in recent lattice QCD work on tetraquark candidates, it has been demonstrated that operators
similar to (\ref{eq:operators}) generate significant overlap to a variety of different four-quark
structures, including mesonic molecules, diquark-antidiquark pairs and two essentially non-interacting
mesons \cite{Alexandrou:2012rm,Abdel-Rehim:2014zwa}.

In previous computations \cite{Wagner:2010ad,Wagner:2011ev,Bicudo:2012qt}, we have considered light
quarks $q \in \{ u,d \}$ (due to technical reasons, the quark mass $m_{u,d}$ was chosen unphysically
heavy corresponding to a pion mass $m_\pi \approx 340 \, \textrm{MeV}$; cf.\ also the first line in
Table~\ref{ta:ensembles}). We studied the scalar isosinglet with antisymmetric spin $j=0$ and flavor $q q
= (ud-du)/\sqrt{2}$ (in the following denoted as the scalar $u/d$ channel), as well as the vector
isotriplet with symmetric spin $j=1$ and flavor $q q \in \{ uu , (ud+du)/\sqrt{2} , dd \}$ (in the
following denoted as the vector $u/d$ channel), which are the two attractive channels between ground
state mesons ($B$ and $B^\ast$). Note that the scalar $u/d$ channel was found to be more attractive than
the vector $u/d$ channel, as expected from quark model considerations (cf.\ eq.\ (\ref{eq:fermibreit})
and the discussion in section~\ref{SEC507}).

In this work, we extend these computations to heavier pairs of light quarks $q q = s s$ and $q q = c c$.
For these symmetric flavor combinations, the only attractive channel for two ground state mesons
($B_{s,c}^{(\ast)}$) is the vector channel, i.e.\ with light quark spin $j = 1$. It corresponds to
$\Gamma = (1 + \gamma_0) \gamma_j$, $j = 1,2,3$, in the creation operator (\ref{eq:operators}).

To be able to study also the scalar channel, i.e.\ $j=0$, with strange and charm quarks, we consider two
hypothetical degenerate flavors with the mass of the strange or the charm quark, which allow to form
antisymmetric flavor combinations $q q = (s^{(1)} s^{(2)} - s^{(2)} s^{(1)}) / \sqrt{2}$ and $q q =
(c^{(1)} c^{(2)} - c^{(2)} c^{(1)}) / \sqrt{2}$. It corresponds to $\Gamma = (1 + \gamma_0) \gamma_5$ in
the creation operator (\ref{eq:operators}).

For further details regarding the lattice QCD computation of $\bar{b} \bar{b}$ potentials, we refer to
\cite{Wagner:2010ad,Wagner:2011ev}. Examples for $q q = (u d - d u) / \sqrt{2}$ (scalar $u/d$ channel)
and for $q q \in \{ uu , (ud+du)/\sqrt{2} , dd \}$ (vector $u/d$ channel) are shown in
\cite{Bicudo:2012qt}, Figure~1.


\subsection{\label{SEC005}Fitting eq.\ (\ref{eq:ansatz}) to lattice QCD $\bar{b} \bar{b}$ potential results}

To describe the lattice QCD $\bar{b} \bar{b}$ potential results $V^\textrm{lat}(r)$ by continuous
functions, we perform uncorrelated $\chi^2$ minimizing fits of eq.\ (\ref{eq:ansatz}), i.e.\ we minimize
\be
\label{EQN004} \chi^2 \ \ = \ \ \sum_{r = r_\textrm{min},\ldots,r_\textrm{max}} \bigg(\frac{V(r) - V^\textrm{lat}(r)}{\Delta V^\textrm{lat}(r)}\bigg)^2
\ee
with respect to the parameters $\alpha$, $d$ and $V_0$, while keeping $p = 2$ fixed (cf.\ the discussion
in section~\ref{SEC388}) \footnote{In principle, one could also use $p$ a a fit parameter. Our lattice
QCD results are, however, not sufficiently precise to extract a stable and precise value also for $p$.
Therefore, we set $p = 2$ as motivated in section~\ref{SEC388}. With this choice, the lattice QCD results
are well described by the fit function~(\ref{eq:ansatz}), i.e.\ the resulting $\chi^2 / \textrm{dof} < 1$
(eq.\ (\ref{EQN004})).}. $\Delta V^\textrm{lat}$ denote the corresponding statistical errors.

We perform these fits for the scalar $u/d$, the vector $u/d$, the scalar $s$, the vector $s$ and the
scalar $c$ channel. The lattice QCD $\bar{b} \bar{b}$ potential of the remaining vector $c$ channel is,
however, strongly screened and consistent with $V^\textrm{lat}(r) = 0$ for $r > 2a$. Such results are not
sufficient to perform a stable fit.

To investigate and quantify systematic errors, we do not only perform a single fit for each of the
mentioned five channels, but a large number of fits, where we vary the following parameters:
\begin{itemize}
\item The range of temporal separations $t_\textrm{min} \leq t \leq t_\textrm{max}$ of the correlation
function $C(t,r)$ (eq.\ (\ref{eq:correlationfunction})) at which $V^\textrm{lat}(r)$ is read off,
according to:
\begin{itemize}
\item $t_\textrm{max} - t_\textrm{min} \geq a$;

\item for $u/d$ channels: \\ $4 a \leq t_\textrm{min}$, $t_\textrm{max} \leq 9 a$;

\item for $s$ and $c$ channels: \\ $10 a \leq t_\textrm{min} \leq 14 a$, $t_\textrm{max} \leq 19 a$
\end{itemize}
(small $t_\textrm{min}$ might lead to a contamination by excited states; large $t_\textrm{min}$ and $t_\textrm{max}$ drastically increase statistical errors).

\item The range of spatial $\bar{b} \bar{b}$ separations $r_\textrm{min} \leq r \leq r_\textrm{max}$
considered in the $\chi^2$ minimizing fit (\ref{EQN004}), according to:
\begin{itemize}
\item for the vector $u/d$ channel: \\ $r_\textrm{min} = 2 a$ \footnote{Our lattice QCD results are not
sufficiently precise to allow stable fits with $r_\textrm{min} = 3 a$ for the vector $u/d$ channel.};

\item for all other channels: \\ $r_\textrm{min} \in \{ 2a , 3a \}$;

\item for $u/d$ channels: \\ $r_\textrm{max} \in \{ 8a , 9a , 10a \}$;

\item for $s$ and $c$ channels: \\ $r_\textrm{max} \in \{ 7a , 8a \}$
\end{itemize}
($V^\textrm{lat}(r)$ at small $r < 2 a$ are expected to suffer from sizable lattice discretization errors, while $V^\textrm{lat}(r)$ at large $r$ is essentially a constant, i.e.\ has little effect on the relevant fit parameters $\alpha$ and $d$).
\end{itemize} 
For each of the fitting parameters $\alpha$, $d$ and $V_0$, we construct a distribution by considering
the results of all the above listed fits weighted by $\exp(-\chi^2/\textrm{dof})$ with $\chi^2$ from eq.\
(\ref{EQN004}). The central values of $\alpha$, $d$ and $V_0$ are then defined as the medians of the
corresponding distributions and the lower/upper systematic uncertainties are given by the difference of
the 16th/84th percentiles to the medians (in the case of a Gaussian distribution, an uncertainty defined
in this way would correspond to its width, i.e.\ $1 \sigma$). Since in general the distributions are
asymmetric, the systematic uncertainties are asymmetric as well. For more details regarding this method
of estimating systematic errors we refer to \cite{Cichy:2012vg}. 

Finally, to include statistical errors, we compute the jackknife errors of the medians of $\alpha$, $d$
and $V_0$ and add them in quadrature to the corresponding systematic uncertainties.

\begin{figure}[htb]
\includegraphics[angle=-90,width=0.8\columnwidth]{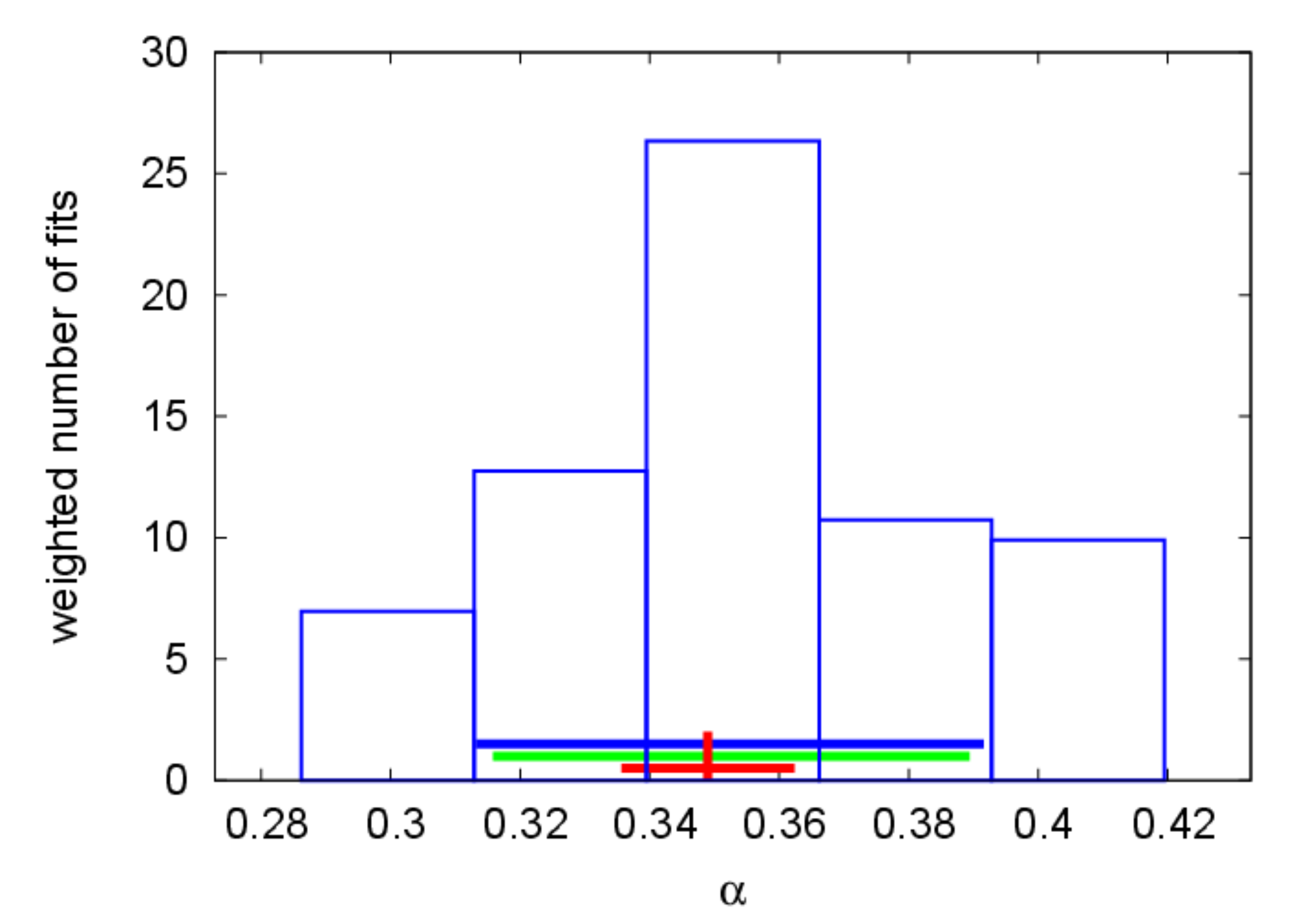}
\includegraphics[angle=-90,width=0.8\columnwidth]{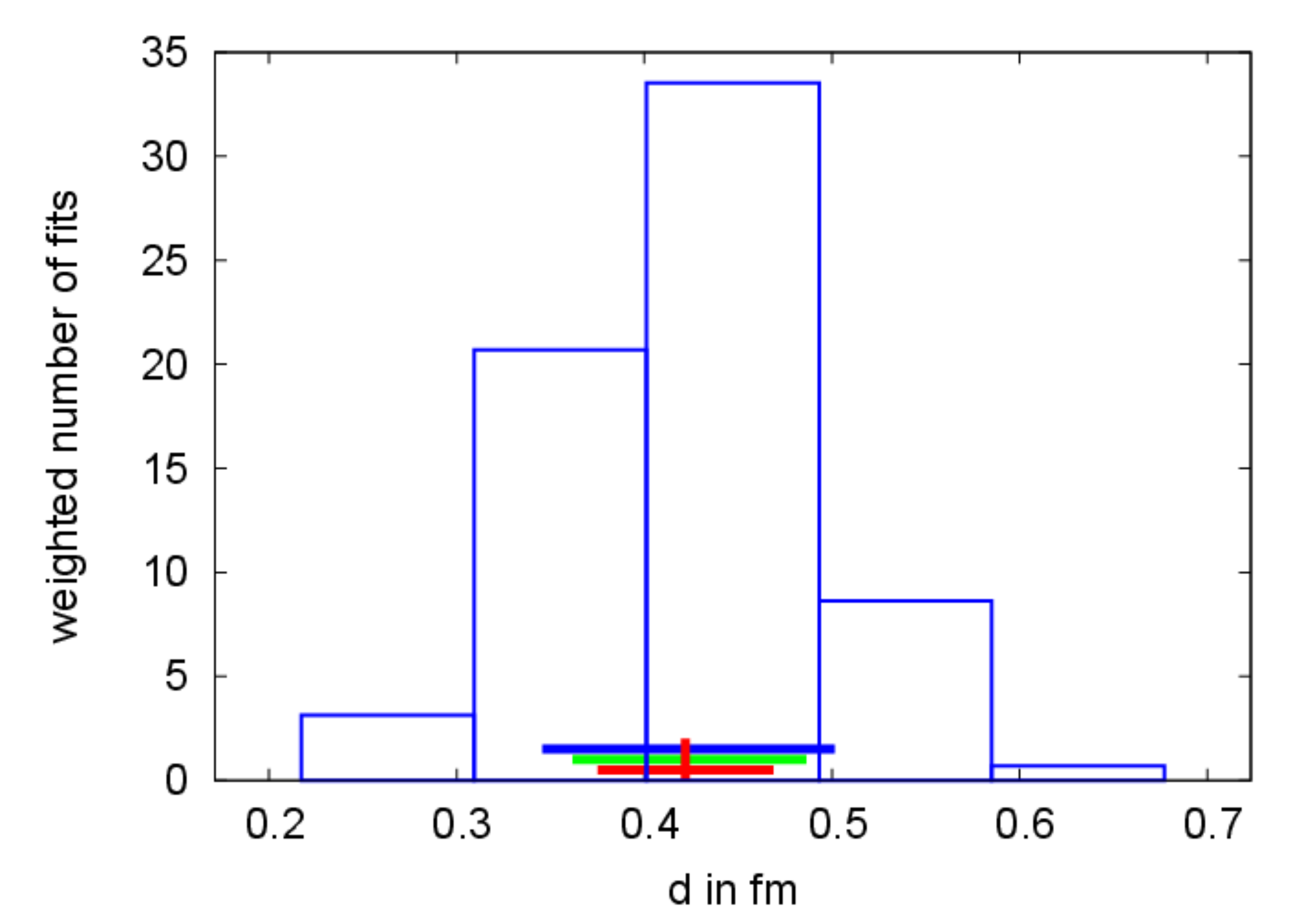}
\caption{\label{FIGhistograms}(Color online). Histograms used to estimate systematic errors for $\alpha$
and $d$ for the scalar $u/d$ channel (green, red and blue bars represent systematic, statistical and
combined errors, respectively).}
\end{figure}

To illustrate this error estimation procedure, we show in Figure \ref{FIGhistograms} example histograms
representing the distribution of $\alpha$ and $d$ for the scalar $u/d$ channel. The green, red and blue
bars correspond to the systematic, statistical and combined errors, respectively. In the following, we
will always use and quote the combined errors represented by the blue bars.

\begin{figure*}[htb]
\includegraphics[width=0.66\columnwidth]{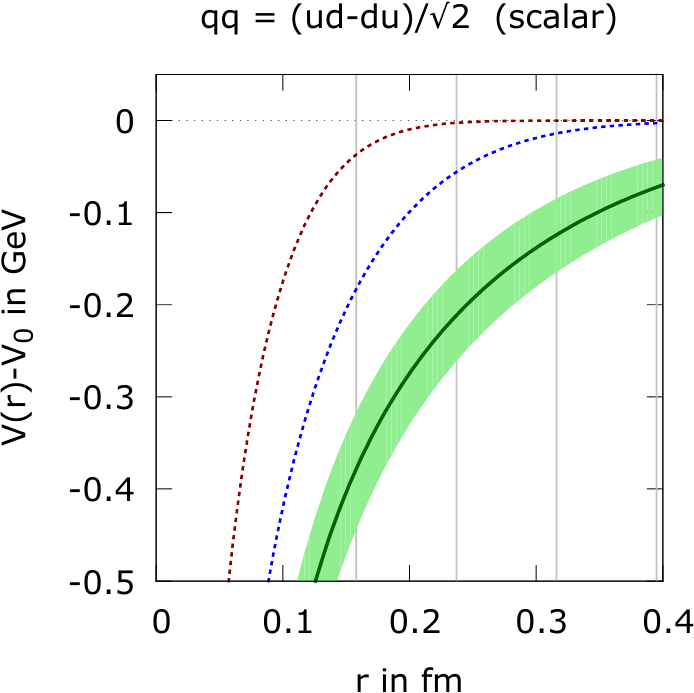}
\includegraphics[width=0.66\columnwidth]{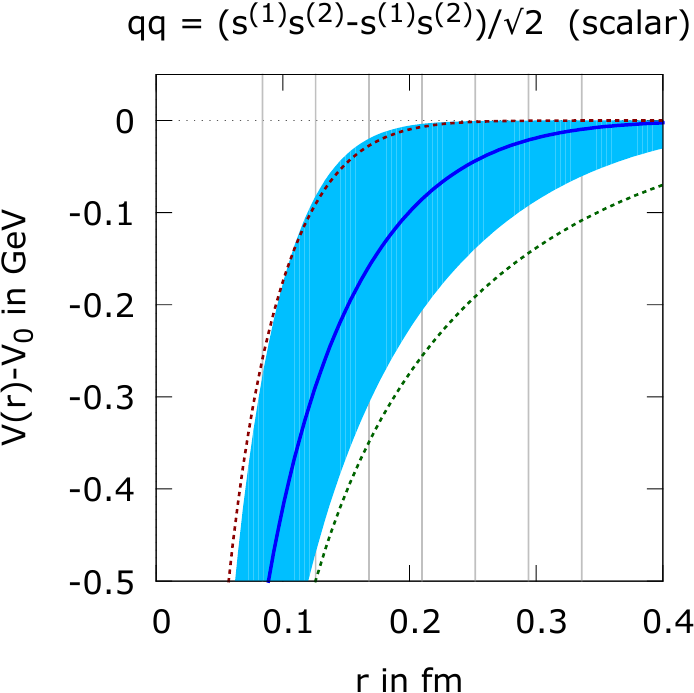}
\includegraphics[width=0.66\columnwidth]{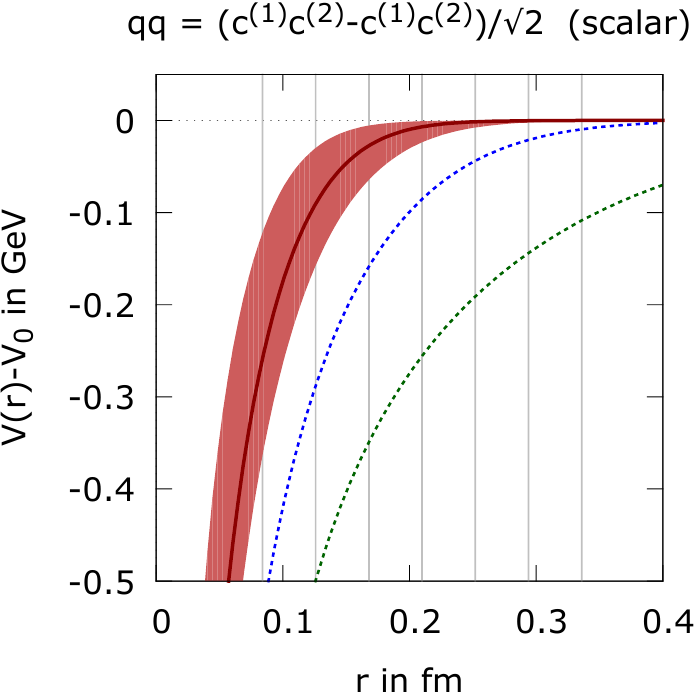} \\
\vspace{0.3cm}
\includegraphics[width=0.66\columnwidth]{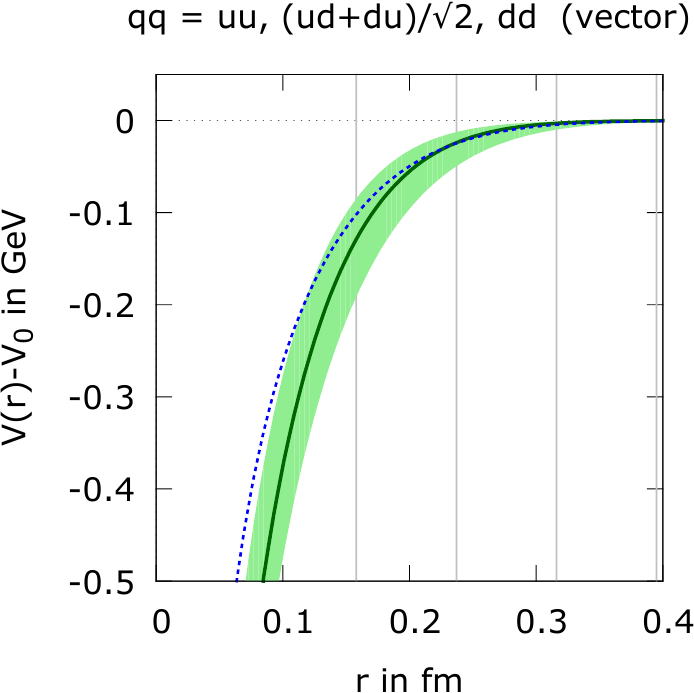}
\includegraphics[width=0.66\columnwidth]{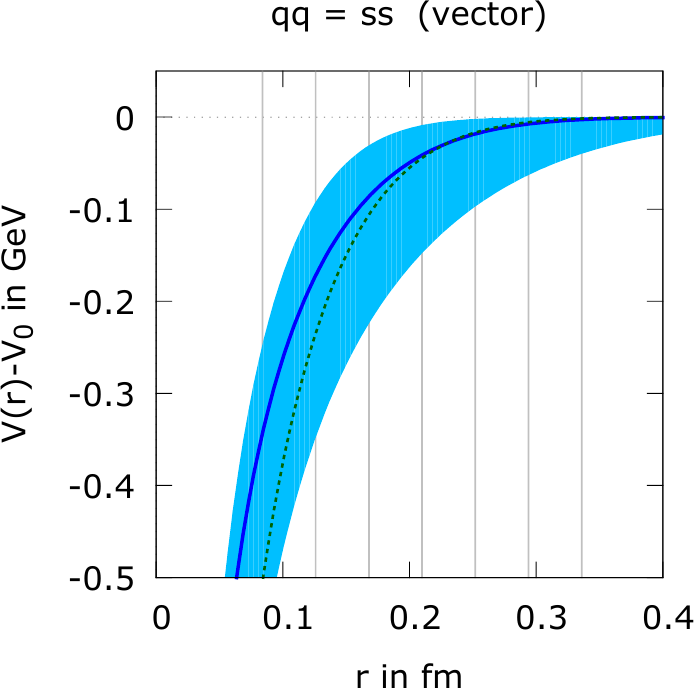}
\phantom{\includegraphics[width=0.66\columnwidth]{figures/potential_scalar_charm_crp.pdf}}
\caption{\label{FIG100}(Color online). $\bar{b} \bar{b}$ potentials in the presence of two lighter quarks $q q$ ($q q$ flavor: up/down in green, strange in blue, charm in red; $q q$ spin: $j=0$, i.e.\ scalar, in the upper line, $j=1$, i.e.\ vector, in the lower line). The plotted curves with the error bands correspond to eq.\ (\ref{eq:ansatz}) with the parameter sets from Table~\ref{ta:fitAEK}. Vertical lines indicate lattice separations $r = 2a,3a,\ldots$ of lattice QCD potential results $V^\textrm{lat}(r)$ used to generate the parameter sets from Table~\ref{ta:fitAEK} via $\chi^2$ minimizing fits.}
\end{figure*}

The final results for $\alpha$ and $d$ are collected in Table~\ref{ta:fitAEK}. Note that within errors
they agree with the model considerations and crude quantitative expectations discussed in
section~\ref{sec:II_model}. We do not list results for $V_0$, since it is an irrelevant constant
corresponding to twice the mass of a static-light meson. The fit function (\ref{eq:ansatz}) with the
parameter sets from Table~\ref{ta:fitAEK} and the corresponding error bands are shown in
Figure~\ref{FIG100}. Clearly, these results confirm the qualitative expectations discussed in
section~\ref{SEC507}:
\begin{itemize}
\item[(1)] The screening of the $\bar{b} \bar{b}$ interaction is stronger for heavier light quarks $q q$.

\item[(2)] The scalar channels are more attractive than the corresponding vector channels.
\end{itemize}

\begin{table}[htb]
\begin{tabular}{c|c|c|c}
\hline
$q q$ & spin & $\alpha$ & $d$ in fm \\
\hline
$(ud-du)/\sqrt{2}$ & scalar & $0.35_{-0.04}^{+0.04}$ & 
$0.42_{-0.08}^{+0.08}$ \\
& & & \vspace{-0.35cm} \\
$uu$, $(ud+du)/\sqrt{2}$, $dd$ & vector & $0.29_{-0.06}^{+0.04}$ & 
$0.16_{-0.01}^{+0.02}$ \\
\hline
$(s^{(1)}s^{(2)}-s^{(2)}s^{(1)})/\sqrt{2}$ & scalar & 
$0.27_{-0.05}^{+0.08}$ & $0.20_{-0.10}^{+0.10}$ \\
& & & \vspace{-0.35cm} \\
$ss$ & vector & $0.18_{-0.02}^{+0.09}$ & $0.18_{-0.05}^{+0.11}$ \\
\hline
$(c^{(1)}c^{(2)}-c^{(2)}c^{(1)})/\sqrt{2}$ & scalar & 
$0.19_{-0.07}^{+0.12}$ & $0.12_{-0.02}^{+0.03}$ \\
\hline
\end{tabular}
\caption{\label{ta:fitAEK}Parameters $\alpha$ and $d$ obtained from $\chi^2$ minimizing fits of (\ref{eq:ansatz}) to lattice QCD $\bar{b} \bar{b}$ potential results.}
\end{table}

\section{\label{sec:IV_binding}Dependence of the existence of $q q \bar{b} \bar{b}$ tetraquark states on the light quark mass}

In \cite{Bicudo:2012qt}, we have found evidence for a bound state in the scalar $u/d$ channel, i.e.\ the
existence of a $q q \bar{b} \bar{b} = u d \bar{b} \bar{b}$ tetraquark. For heavier quarks $q q$, the
effective $\bar{b} \bar{b}$ potentials are less attractive. This has qualitatively been anticipated in
section~\ref{sec:II_model} and quantified in section~\ref{sec:III_latticefit} (in particular cf.\ the
resulting values for $\alpha$ and $d$ in Table~\ref{ta:fitAEK} and the plots in Figure~\ref{FIG100}).
Thus, for a sufficiently heavy pair of light quarks $q q$ we expect that the $q q \bar{b} \bar{b}$ system
will not anymore be able to form a bound state. In the following, we investigate whether this is already
the case for strange and/or charm quark masses. We also study the vector channels.


\subsection{\label{SEC800}The $\bar{b} \bar{b}$ Hamiltonian}

We define $U(r) = V(r)|_{V_0 = 0, p = 2}$ with $V(r)$ from of eq.\ (\ref{eq:ansatz}). $U(r)$ with a set
of fit parameters $\alpha$ and $d$ from Table~\ref{ta:fitAEK} corresponds to the ground state energy of a
$q q \bar{b} \bar{b}$ 4-quark system in a specific channel minus the energy of a pair of far separated
$B_{(s,c)}^{(\ast)}$ mesons. Thus, the corresponding Hamiltonian for the relative coordinate of the
$\bar{b} \bar{b}$ quarks is
\be
\label{eq:schro} H \ \ = \ \ \frac{\mathbf{p}^2}{2 \mu} + 2 m_H + U(r),
\ee
where $\mu = m_H/2$ is the reduced mass. At large separations, each $\bar{b}$ quark carries the mass of a
$B_{(s,c)}^{(\ast)}$ meson because of screening, and thus $m_H = m_{B_{(s,c)}^{(\ast)}}$. At small
separations, $m_H = m_b$ could be more appropriate. Throughout this section, we always consider two
choices, $m_H = m_{B_{(s,c)}}$ and $m_H = m_b$, which yield qualitatively identical results. Note that
any dependence on the heavy $\bar{b}$ spins is neglected, because $V(r)$ has been computed in the static
limit $m_b \rightarrow \infty$. Since the $\bar{b}$ quarks are quite heavy, we expect the static limit to
be a reasonable approximation.

In classical mechanics, the $\bar{b} \bar{b}$ separation $r$ would vanish for the ground state, but after
quantizing the system, a bound 4-quark state ($E < 2 m_H$) may not exist anymore.


\subsection{An analytical estimate for $q q \bar{b} \bar{b}$ binding}

In \cite{Bicudo:2012qt}, we have derived an approximate analytical rule for the existence/non-existence
of a bound $q q \bar{b} \bar{b}$ state using the Bohr-Sommerfeld quantization condition. If
\be
\label{condition} \mu \alpha d \ \ \geq \ \ \frac{9 \pi^2}{128 \times 2^{1/p} (\Gamma(1 + 1/2 p))^2}
\ee
is fulfilled, there should be at least one bound state. The right hand side of this rule has a rather
moderate dependence on the exponent $p$. For example, when $p$ increases from the expected values of
$1.5$ to $2.0$ (cf.\ section~\ref{SEC388}), the right hand side only changes from $0.55$ to $0.60$. Thus,
the existence of a bound state mainly depends on the product of parameters $\mu \alpha d$. 

With the medians for the parameters $\alpha$ and $d$ (cf.\ Table~\ref{ta:fitAEK}), we determine the left
hand side of eq.\ (\ref{condition}). For the reduced mass, we use both $m_H = m_{B_{(s,c)}}$ ($m_B = 5279
\, \textrm{MeV}$, $m_{B_s} = 5367 \, \textrm{MeV}$, $m_{B_c} = 6276 \, \textrm{MeV}$ \cite{PDG}), which
is certainly a good choice for large $\bar{b} \bar{b}$ separations, and $\mu = m_b/2$ ($m_b = 4977 \,
\textrm{MeV}$, from quark models \cite{Godfrey:1985xj}), which might be more appropriate for small
$\bar{b} \bar{b}$ separations (cf.\ the discussion in section~\ref{SEC800}). The results for $\mu \alpha
d$ for the $u/d$, $s$ and scalar $c$ and vector channels are collected in Table~\ref{ta:pedros_eq}. For
the scalar $u/d$ channel, there is strong indication for the existence of a tetraquark (i.e.\ $\mu \alpha
d \gg 0.60$), which confirms our findings from \cite{Bicudo:2012qt}. For the vector $s$ channel and for
charm quarks, bound $q q \bar{b} \bar{b}$ states are not expected (i.e.\ $\mu \alpha d \ll 0.60$). For
the vector $u/d$ and the scalar $s$ channel the situation is less clear. A more rigorous and quantitative
analysis is needed, which is part of the following section.

\begin{table}[htb]
\begin{tabular}{c|c|c|c}
\hline
\multicolumn{4}{c}{$\mu \alpha d$} \\
\hline
$q q$                                      & spin   & $m_H = m_{B_{(s,c)}}$ & $m_H = m_b$ \\
\hline
$(ud-du)/\sqrt{2}$                         & scalar & $1.97$ & $1.86$ \\
$uu$, $(ud+du)/\sqrt{2}$, $dd$             & vector & $0.60$ & $0.57$ \\
\hline
$(s^{(1)}s^{(2)}-s^{(2)}s^{(1)})/\sqrt{2}$ & scalar & $0.74$ & $0.69$ \\
$ss$                                       & vector & $0.44$ & $0.41$ \\
\hline
$(c^{(1)}c^{(2)}-c^{(2)}c^{(1)})/\sqrt{2}$ & scalar & $0.34$ & $0.27$ \\
\hline
\end{tabular}
\caption{\label{ta:pedros_eq}Values for $\mu \alpha d$, which represent the left hand side of eq.\
(\ref{condition}). Values $> 0.60$ (right hand side of eq.\ (\ref{condition}) for $p=2$) point towards
the existence of a bound $q q \bar{b} \bar{b}$ state, while values $< 0.60$ are an indication against the
existence of such a state.}
\end{table}


\subsection{Numerical solution of the Schr\"odinger equation}

To investigate the existence of a bound state more rigorously, we solve the Schr\"odinger equation with
the Hamiltonian (\ref{eq:schro}) numerically. The strongest binding is expected in an s-wave, for which
the radial equation is 
\be
\label{EQN002} \bigg(-\frac{1}{2 \mu} \frac{d^2}{dr^2} + U(r)\bigg) R(r) \ \ = \ \ \underbrace{\Big(E - 2 m_H \Big)}_{= E_B} R(r)
\ee
with the wave function $\psi = \psi(r) = R(r) / r$. If $E_B = E - 2 m_H < 0$, $-E_B$ can be interpreted
as the binding energy. We proceed as explained in \cite{Bicudo:2012qt} and solve this equation by
imposing Dirichlet boundary conditions $R(r = \infty) = 0$ and using 4th order Runge-Kutta shooting.

For the scalar $u/d$ channel, the lowest eigenvalue $E_B < 0$, which implies the existence of a bound
four-quark state. For all other channels, i.e.\ the vector $u/d$ and the $s$ and $c$ channels, $E_B > 0$,
i.e.\ the corresponding $q q \bar{b} \bar{b}$ tetraquarks will most likely not exist in these channels
\footnote{As mentioned previously in section~\ref{SEC005}, the lattice QCD results for the vector $c$
channel are not sufficient to perform a quantitative analysis. The $\bar{b} \bar{b}$ potential in this
channel is, however, much less attractive than in the other channels, e.g.\ the scalar $c$ channel.
Therefore, a bound four-quark state in the vector $c$ channel can be excluded.}. These findings confirm
the analytical estimates obtained in the previous subsection (eq.\ (\ref{condition}) and
Table~\ref{ta:pedros_eq}).

\begin{figure}[htb]
\includegraphics[angle=-90,width=0.8\columnwidth]{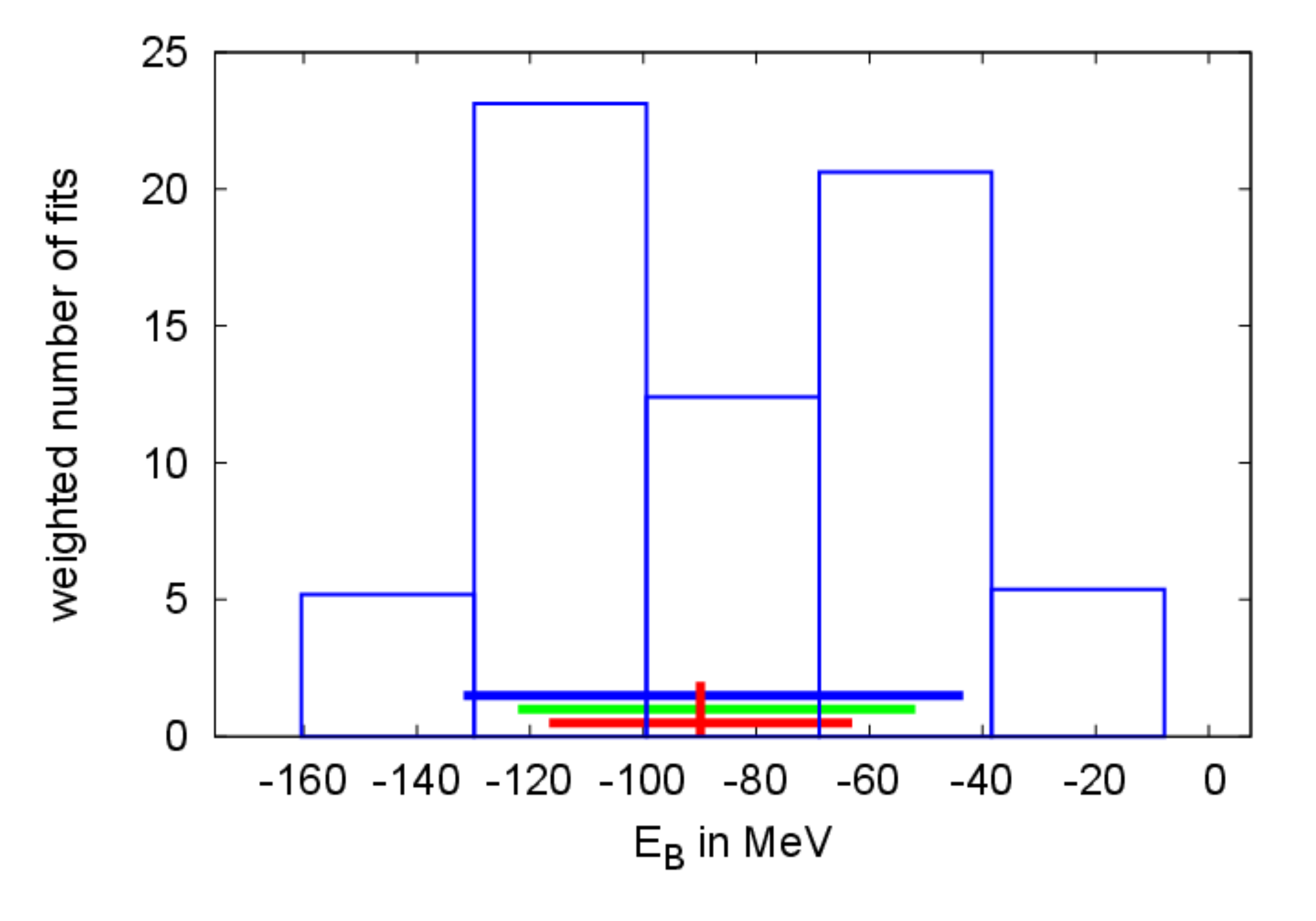}
\caption{\label{FIGhistogramsE}(Color online). Histogram used to estimate the systematic error for the
binding energy $E_B$ for the scalar $u/d$ channel and $m_H = m_B$ (green, red and blue bars represent
systematic, statistical and combined errors, respectively).}
\end{figure}

The central value and the combined systematic and statistical error for the binding energy $E_B$ of the
tetraquark state in the scalar $u/d$ channel is obtained by the method discussed in section~\ref{SEC005}
(generating a distribution for $E_B$ from the fits listed in section~\ref{SEC005}):
\begin{eqnarray}
 & & \hspace{-0.7cm} E_B = -90_{-42}^{+46} \, \textrm{MeV} \quad \textrm{(for }m_H = m_B\textrm{)}, \\
 & & \hspace{-0.7cm} E_B = -93_{-43}^{+47} \, \textrm{MeV} \quad \textrm{(for }m_H = m_b\textrm{)} .
\end{eqnarray}
These binding energies are roughly twice as large as their combined systematic and statistical errors. In
other words, the confidence level for this $u d \bar{b} \bar{b}$ tetraquark state is around $2 \sigma$.
The corresponding histogram for $m_H = m_B$ is shown in Figure~\ref{FIGhistogramsE}.

To crudely quantify also the non-existence of bound four-quark states in the remaining channels, we
determine numerically by which factors the heavy masses $m_H$ in the Schr\"odinger equation
(\ref{EQN002}) have to be increased to obtain bound states, i.e.\ tiny but negative energies $E_B$ (the
potentials $U(r)$ are kept unchanged, i.e.\ we stick to the medians for $\alpha$ and $d$ from
Table~\ref{ta:fitAEK}). The resulting factors are collected in Table~\ref{ta:binding}. While the scalar
$s$ channel is quite close to be able to host a bound state, the scalar $c$ channel and the vector
channels are rather far away, since they would require $\bar{b}$ quarks approximately $1.6 \ldots 3.3$
times as heavy as they are in nature. Note that the factors listed in Table~\ref{ta:binding} could also
be relevant for quark models aiming at studying the binding of tetraquarks quantitatively.

\begin{table}[htb]
\begin{tabular}{c|c|c|c}
\hline
$q q$                                      & spin   & $m_H = m_{B_{(s,c)}}$ & $m_H = m_b$ \\
\hline
$(ud-du)/\sqrt{2}$                         & scalar & $0.46$ & $0.49$ \\
$uu$, $(ud+du)/\sqrt{2}$, $dd$             & vector & $1.49$ & $1.57$ \\
\hline
$(s^{(1)}s^{(2)}-s^{(2)}s^{(1)})/\sqrt{2}$ & scalar & $1.20$ & $1.29$ \\
$ss$                                       & vector & $2.01$ & $2.18$ \\
\hline
$(c^{(1)}c^{(2)}-c^{(2)}c^{(1)})/\sqrt{2}$ & scalar & $2.57$ & $3.24$ \\
\hline
\end{tabular}
\caption{\label{ta:binding}Factors by which the mass $m_H$ has to be multiplied to obtain a tiny but
negative energy $E_B$. Factors $\ll 1$ indicate strongly bound states, while for values $\gg 1$ bound
states are essentially excluded.}
\end{table}

\begin{figure}[htb]
\includegraphics[width=\columnwidth]{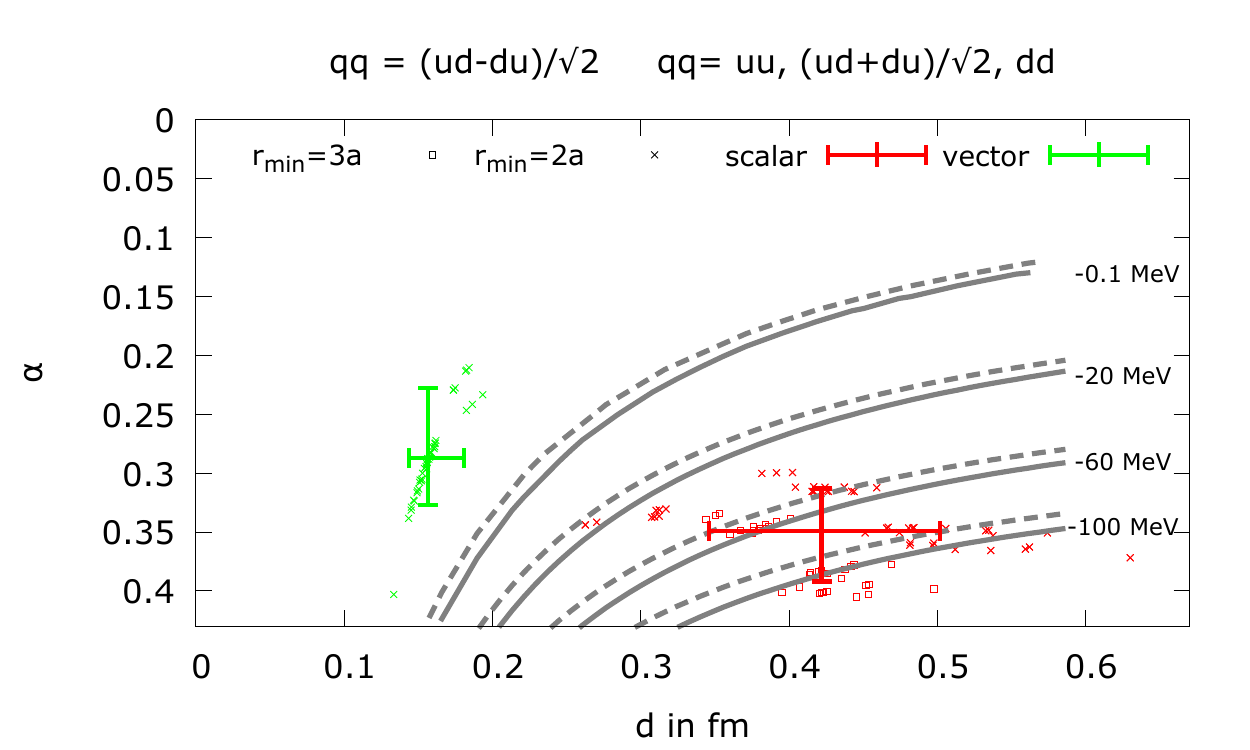} \\
\includegraphics[width=\columnwidth]{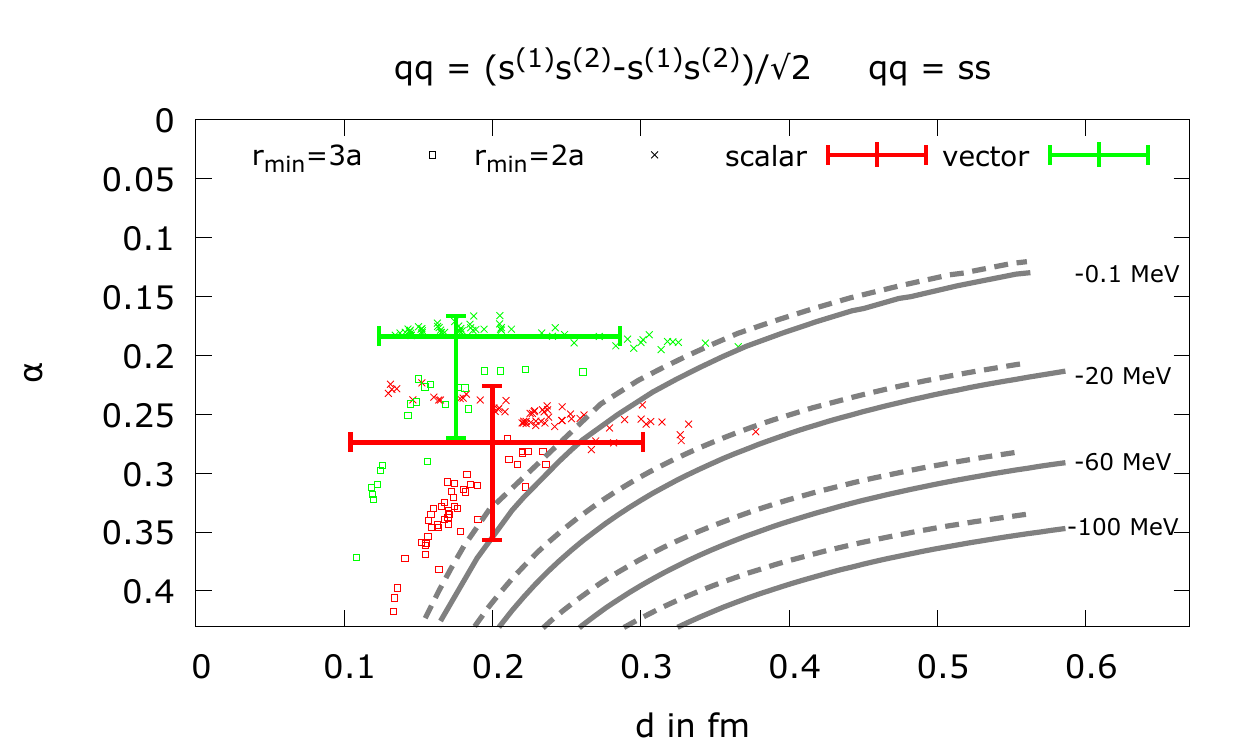} \\
\includegraphics[width=\columnwidth]{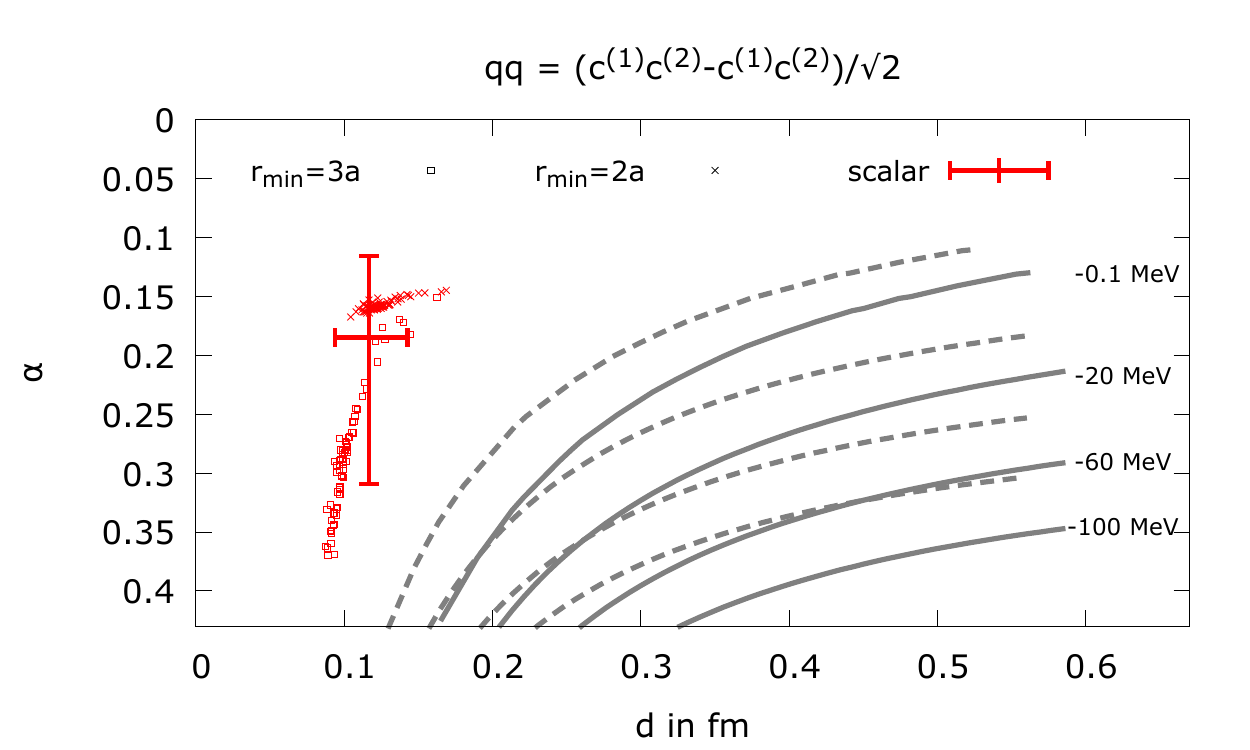}
\caption{\label{FIG340}(Color online). Binding energy isolines $E_B(\alpha,d) = \textrm{constant}$ in
the $\alpha$-$d$-plane for $u/d$, $s$ and $c$ light quarks $q q$, respectively (gray dashed lines: $m_H =
m_{B_{(s,c)}}$; gray solid lines: $m_H = m_b$). The red and green dots represent the fits of eq.\
(\ref{eq:ansatz}) to lattice QCD $\bar{b} \bar{b}$ potential results, while the red and green bars are
the corresponding combined systematic and statistical errors.}
\end{figure}

In Figure~\ref{FIG340}, we present our results in an alternative graphical way. Binding energy isolines
$E_B(\alpha,d) = \textrm{constant}$ are plotted in the $\alpha$-$d$-plane starting at a tiny energy $E_B
= -0.1 \, \textrm{MeV}$ up to rather strong binding, $E_B = -100 \, \textrm{MeV}$ (gray dashed lines have
been computed with $m_H = m_{B_{(s,c)}}$, gray solid lines with $m_H = m_b$). The three plots correspond
to $u/d$, $s$ and $c$ light quarks $q q$, respectively. Each fit of eq.\ (\ref{eq:ansatz}) to lattice QCD
$\bar{b} \bar{b}$ potential results (cf.\ the detailed discussion about systematic error estimation for
$\alpha$ and $d$ in section~\ref{SEC005}) is represented by a dot (red: scalar channels; green: vector
channels; crosses: $r_\textrm{min} = 2a$; boxes: $r_\textrm{min} = 3a$). The extensions of these point
clouds represent the systematic uncertainties with respect to $\alpha$ and $d$. If a point cloud is
localized above or left of the isoline with $E_B = -0.1 \, \textrm{MeV}$ (approximately the binding
threshold), the corresponding four quarks $q q \bar{b} \bar{b}$ will not form a bound state. A
localization below or right of that isoline is a strong indication for the existence of a tetraquark. In
case the point cloud is intersected by that isoline, the estimated systematic error is too large to make
a definite statement regarding the existence or non-existence of a bound four-quark state. The big red
and green bars in horizontal and vertical direction represent the combined systematic and statistical
errors of $\alpha$ and $d$, as quoted in Table~\ref{ta:fitAEK}. One can observe and conclude the
following from Figure~\ref{FIG340}:
\begin{itemize}
\item There is clear evidence for a tetraquark state in the scalar $u/d$ channel.

\item The scalar $s$ channel is close to binding/unbinding. A definite statement with our currently available lattice QCD data is not possible.

\item the scalar $c$ and all vector channels do not host a bound four-quark state.
\end{itemize}
These findings are consistent with the results presented above in Table~\ref{ta:pedros_eq} and Table~\ref{ta:binding}.

\section{\label{sec:V_conclusion}Conclusions and outlook}

In a previous publication \cite{Bicudo:2012qt}, we have found indication for the existence of a $q q
\bar{b} \bar{b}$ tetraquark with $q q = (ud-du) / \sqrt{2}$ (i.e.\ in the scalar $u/d$ channel). In this
work, we have extended these studies by considering for $q q$ not only $u/d$, but also heavier $s$ and
$c$ quarks. In contrast to \cite{Bicudo:2012qt}, we have also investigated and quantified systematic
uncertainties in detail.

Our main results are the following:
\begin{itemize}
\item We confirm the $u d \bar{b} \bar{b}$ tetraquark state in the scalar $u/d$ channel predicted in
our previous paper \cite{Bicudo:2012qt} with confidence level $\approx 2 \sigma$. The overall quantum
numbers of this state are $I(J^P) = 0(1^+)$.

\item There is no bound four-quark state in the vector $u/d$ channel ($I(J^P) =\in \{ 1(0^+) , 1(1^+) ,
1(2^+) \}$). Note, however, that we have been using unphysically heavy $u/d$ quarks ($m_\pi \approx 340
\, \textrm{MeV}$). Since decreasing the light quark mass should enhance binding, it will be interesting
to explore in the future whether a bound four-quark state exists at physically light $u/d$ quark mass.

\item $s s \bar{b} \bar{b}$ and $c c \bar{b} \bar{b}$ tetraquarks, which correspond to the vector $s$
and $c$ channels ($J^P \in \{ 0^+ , 1^+ , 2^+ \}$), do not exist.

\item It is of conceptual interest to introduce a hypothetical second $s$ or $c$ quark flavor. Then it
is possible to also study the scalar $s$ and $c$ channels, i.e.\ $((s^{(1)} s^{(2)} - s^{(2)} s^{(1)}) /
\sqrt{2}) \bar{b} \bar{b}$ and $((c^{(1)} c^{(2)} - c^{(2)} c^{(1)}) / \sqrt{2}) \bar{b} \bar{b}$ systems
($J^P = 1^+$). While in the scalar $c$ channel there is no bound four-quark state, the situation is less
clear for $s$ quarks. Improved lattice QCD results (less statistical errors, finer resolution of $\bar{b}
\bar{b}$ separations) are needed before a definite statement can be made. Binding in the hypothetical
scalar $s$ channel would indicate a fortiori binding for four-quark systems $((us - su)/ \sqrt{2})
\bar{b} \bar{b}$ and $((ds - sd)/ \sqrt{2}) \bar{b} \bar{b}$. Such light-strange channels would then be
highly relevant for experimental tetraquark searches.
\end{itemize}
We consider these results to be important because they indicate both to experimental collaborations and
to quark model phenomenologists which $q q \bar{b} \bar{b}$ tetraquarks are expected to exist and which
are not.

To supply data for future quark model studies of tetraquarks, we also provide parameterizations of the
potential of two static antiquarks $\bar{b} \bar{b}$ in the presence of two lighter quarks $q q$, where
$q q \in \{ (ud-du)/\sqrt{2} \ , \ uu, (ud+du)/\sqrt{2}, dd \ , \
(s^{(1)}s^{(2)}-s^{(2)}s^{(1)})/\sqrt{2} \ , \ ss \ , \ (c^{(1)}c^{(2)}-c^{(2)}c^{(1)})/\sqrt{2} \}$.
Moreover, we have determined quantitatively for these channels by which factor the heavy quark or meson
mass $m_H$ has to be increased to obtain a tetraquark state.

It is also interesting to compare our findings to other groups
studying the same or similar systems using, however, different
theoretical approaches.
For instance in \cite{Du:2012wp}, in the framework of QCD sum rules,
binding for flavors equivalent to $u d \bar b \bar b$, $u s \bar b
\bar b$ and  $s s \bar b \bar b$ has been found, and no binding for
doubly charmed tetraquarks. However, these bound systems have $J^P=
0^-$ and  $J^P= 1^-$ different from our results.
Another example using the Dyson-Schwinger framework is
\cite{Heupel:2012ua}, where a tetraquark composed of four charm
quarks, i.e.\ $c c \bar{c} \bar{c}$, has recently been predicted with
a mass significantly lighter than $2 m_{\eta_c}$. In principle our
static antiquarks can also be considered as a crude approximation of
$\bar{c} \bar{c}$. Since we do not find a bound state for $q q = c c$,
there seems to be a qualitative discrepancy to our results, which would be interesting to understand and to resolve.

As an outlook, it would be interesting to decrease the light $u/d$ quark mass to their physical value,
since this should increase the radius of a $B$ meson, reduce screening and, therefore, lead to a larger
binding energy. As mentioned above, a tetraquark could then also exist in the vector $u/d$ channel.
Additionally, lighter $u/d$ quark masses may also allow the study of light meson exchange interactions
between the two $B$ mesons. Because simulations and computations at lighter $u/d$ quark masses are
computationally very expensive, we leave them for a future publication.

Since there is a bound state for $q q = (ud-du)/\sqrt{2}$, and possibly even for $q q =
(s^{(1)}s^{(2)}-s^{(2)}s^{(1)})/\sqrt{2}$, it will be very interesting to investigate $u s \bar{b}
\bar{b}$ (or equivalently $d s \bar{b} \bar{b}$) systems. This will, however, require additional
computations and also the implementation of certain modifications in our analysis procedure. We plan to
study such flavor combinations in the near future.

Another interesting, but very challenging task, is to include corrections due to the heavy $\bar{b}
\bar{b}$ spins. While in principle it is possible to compute such corrections using lattice QCD (cf.\
\cite{Koma:2006si,Koma:2006fw}, where this has been pioneered for the standard static quark-antiquark
potential), in practice we expect this to be extremely hard for $q q \bar{b} \bar{b}$ systems.
Therefore, a more promising and realistic approach seems to include such spin-dependent interactions in
the Schr\"odinger equation, which will result in a coupled channel differential equation. We are
currently in the process of exploring this approach, where first promising qualitative results have
recently been presented at a conference \cite{Jonas2015}.

Once these techniques are fully developed for $q q \bar{b} \bar{b}$ systems, it will be most interesting
to extend them to $q \bar{q} b \bar{b}$ systems and to study the crypto-exotic $b \bar{b}$ tetraquark
candidates observed by the BELLE collaboration \cite{Belle:2011aa}.


\acknowledgments

P.B.\ thanks IFT for hospitality and CFTP, grant FCT UID/FIS/00777/2013, for support. M.W.\ and A.P.\ acknowledge support by the Emmy Noether Programme of the DFG (German Research Foundation), grant WA 3000/1-1.

This work was supported in part by the Helmholtz International Center for FAIR within the framework of the LOEWE program launched by the State of Hesse.

Calculations on the LOEWE-CSC high-performance computer of Johann Wolfgang Goethe-University Frankfurt am Main were conducted for this research. We would like to thank HPC-Hessen, funded by the State Ministry of Higher Education, Research and the Arts, for programming advice.



\end{document}